\documentclass[12pt]{article}
\def\diy{\displaystyle}
\def\C{\mathbb C}

\def\E{\mathbb E}

\def\N{\mathbb N}
\def\P{\mathbb P}
\def\R{\mathbb R}

\def\Z{\mathbb Z}

\def\ol{\overline}

\def\wh{\widehat}

\def\cA{{\mathcal A}}

\def\cD{{\mathcal D}}

\def\cJ{{\mathcal J}}

\def\bfE{{\mathbf E}}
\def\dist{{\rm{dist}}}

\def\u0{{\underline 0}}
\def\uX{{\underline X}}
\def\uu{{\underline u}}
\def\uv{{\underline v}}

\def\ux{{\underline x}}
\def\uy{{\underline y}}
\def\uz{{\underline z}}
\def\wh{\widehat}

\def\om{{\omega}}
\def\Lam{{\Lambda}}

\def\lam{{\lambda}}
\def\half{\frac{1}{2}}
\def\pt{\partial}
\def\pn{\par\noindent}
\def\pmn{\par\medskip\noindent}
\def\psn{\par\smallskip\noindent}
\def\z2{{\Z^2}}
\def\zp2{{\Z^2_{\geq}}}

\def\myset#1{{\left\{\,#1\,\right\}}}
\def\esm#1{{\E\left[ \, #1 \, \right]}}
\def\esms#1#2{{\E_{#1}\left[ \, #2 \, \right]}}
\def\abs#1{{\left| \, #1 \, \right| }}

\def\pr#1{{  \P\left\{ \, #1 \, \right\}  }}

\def\dist{{\,{\rm dist}}}
\def\dinf{{\,{\rm d_\infty}}}

\def\abs#1{{\left| \, #1 \, \right|}}
\def\norm#1{{\left\| \, #1 \, \right\|}}

\def\sq{{{sub-square }}}

\def\myproof#1{{\pmn{\bf Proof of #1: }}}
\def\myrem{{\pmn{\bf Remark. }}}

\def\tto#1{\smash{\mathop{\,\,\,\, \longrightarrow \,\,\,\, }\limits_{#1}}}

\def\vbeta{{1/2}}

\def\vp{{6}}
\def\vq{{24}}
\def\valpha{{3/2}}

\def\vn{{ 6 }}

\def\DSk{{\bf DS$(k,I)$ }}
\def\DSkone{{\bf DS$(k+1,I)$ }}

\usepackage{graphics}

\usepackage{latexsym}
\usepackage{amsfonts, amssymb}
\usepackage{amsmath}
\usepackage{amsthm}
\usepackage{psfrag}
\usepackage{pstricks,pst-node,pst-tree}
\usepackage{graphicx}
\usepackage{array}
\usepackage{rotating}

\newtheorem{Thm}{Theorem}[section]{\bf}{\it}
\newtheorem{Lem}{Lemma}[section]{\bf}{\it}
\newtheorem{Def}{Definition}[section]{\bf}{\it}
{\bf}{\it}

\begin{document}
\title{ Anderson localisation for
an interacting two-particle quantum system on ${\mathbb Z}$}
\author{Victor Chulaevsky $^1$,
Yuri Suhov $^2$
}                     

\date{}

%
%
\maketitle
\pmn
\begin{center}
$^1$
D\'{e}partement de Math\'{e}matiques et Informatique, \\
Universit\'{e} de Reims, Moulin de la Housse, B.P. 1039, \\
51687 Reims Cedex 2, France \\
E-mail: victor.tchoulaevski@univ-reims.fr
\pmn
$^2$ Department of Pure Mathematics and Mathematical Statistics, \\
University of Cambridge, Wilberforce Road, \\
Cambidge CB3 0WB, UK\\
E-mail: Y.M.Suhov@statslab.cam.ac.uk
\end{center}
\newpage
\begin{abstract}
We study spectral properties of a system of two quantum particles on an integer lattice
$\Z$ with a bounded short-range two-body interaction, in an external random potential
field $x\mapsto V(x,\omega )$ with independent, identically distributed values. The main
result is that if the common probability density $f$ of random variables $V(x,\;\cdot\;)$
is analytic in a strip around the real line and the amplitude constant
$g$ is large enough (i.e. the system is at high disorder), then, with
probability one, the spectrum of the two-particle lattice Schr\"{o}dinger operator $H(\om)$
(bosonic or fermionic) is pure point, and all eigen-functions decay exponentially. The
proof given in this paper is based on a refinement of a multiscale analysis (MSA) scheme
proposed by von Dreifus and Klein (\cite{vDK1}), adapted to incorporate lattice systems
with interaction.
\end{abstract}
\section{Introduction}
\label{intro}
\subsection{Random operators}
\label{ssec:1.1}

Random self-adjoint operators appear in various problems of physical origin, in
particular, in solid state physics. For example, they model properties of an ideal or
non-ideal crystal where immovable atoms create an external potential field for moving
electrons. Typically, it is difficult to analyse spectral properties of each sample
operator $H(\omega)$. However, one rarely, if ever, needs such a detailed information. A
more subtle approach is to consider {\it almost every} operator and establish properties
held with probability one. Perhaps the most popular model of a random operator is a
lattice Schr\"{o}dinger operator (LSO) $H(=H^{(1)}_{V,g}(\omega ))$ with a random external
potential. Operator $H$ has the form $H^0+gV$ and acts on functions $f$ from
$\ell_2(\Z^d)$:
$$H\phi (\ux )=H^0\phi (\ux )+gV\phi (\ux )=
\sum_{\langle \uy ,\ux \rangle}f(\uy )+
gV(\ux ,\omega)f(\ux ),\;\;\ux \in\Z^d.\eqno (1.1)$$ Here $H^0$ stands for the kinetic
energy operator (the lattice Laplacian) and $V$ for the potential energy operator.
Further,
$\langle \uy ,\ux \rangle$ indicates a nearest-neighbor
pair of lattice sites $\uy ,\ux \in\Z^d$. Finally, $g$ is an amplitude constant.

In particular, the Anderson model is where $V(\ux ,\omega )$, $\ux \in\Z$, are
real-valued independent, identically distributed (IID) random variables (RVs). This model
describes the motion of a single lattice electron in a potential field generated by
random `impurities' present at sites
$\ux $ of the cubic lattice $\Z^d$ independently for different sites. The
question here is about the character of the spectrum of LSO $H$ in (1.1).

The single-particle Anderson model generated a substantial literature, and Anderson's
localisation in a single-particle system is now well understood. The initial result was
suggested by Sinai in the mid-70's and proved in
\cite{GMP} for one-dimensional case ($d=1$). We refer the reader to
subsequent works \cite{FS}, \cite{MS}, \cite{FMSS},
\cite{vD}, and particularly \cite{vDK1}. A multi-scale analysis (MSA)
scheme proposed in \cite{vDK1} proved to be very general and flexible and has been
applied to different models of disordered media. The scheme was re-fined in \cite{AM} and
\cite{ASFH}. The general result of these papers is that for the Anderson model in any
dimension $d\geq 1$, with a fairly general distribution of $V(\;\cdot\;,\omega )$ and a
sufficiently large amplitude $|g|$, operator $H_{V,g}$ has with probability one a pure
point spectrum, and all its eigen-functions (EFs) decay exponentially fast at infinity
("exponentially localised", in physical terminology). This phenomenon is often called
Anderson, or exponential, localisation.

\subsection{Interacting systems}

    This paper considers a {\it two-particle} Anderson system on a
one-dimensional lattice $\Z$, with {\it interaction}, in a random external potential. The
Hamiltonian/LSO $H\left(=H^{(2)}_{U,V,g}
(\omega )\right)$ is of the form $H^0+U+g(V_1+V_2)$:
$$\begin{array}{l}
H\phi (\ux)=H^0\phi (\ux)+
\left[\big(U+gV_1+gV_2\big)\phi\right](\ux )\\
\;\;=\phi (x_1-1,x_2)+\phi (x_1+1,x_2)+
\phi (x_1,x_2-1)+\phi (x_1,x_2+1) \\
\;\;+\left[U(\ux )+gV(x_1,\omega )+gV(x_2,\omega )\right]
\phi (\ux ),\;\;\ux=(x_1,x_2)
\in\z2_{\geq}.\end{array} \eqno(1.2)$$
Here, as before, $H^0$ stands for the kinetic energy operator (the lattice Laplacian),
and $U+gV_1+ gV_2$ is the potential energy operator; all operators act in the
two-particle Hilbert space
$\ell^2(\z2_{\geq})$. Next, $\z2_{\geq}$ is the 'sub-diagonal half'
of the two-dimensional lattice $\Z^2$:
$$\z2_{\geq}=\{\ux =(x_1,x_2):\;x_1,x_2\in\Z,\;x_1\geq x_2\}.
\eqno(1.3)$$
A boundary condition on the diagonal
$$\partial\Z^2_{\geq}=\{\ux=(x_1,x_2):\;x_1=x_2\}$$
specifies the statistics of the two-particle system: it is a reflection condition for a
bosonic and zero (Dirichlet's) condition for a fermionic system. Consequently, in the RHS
of (1.2), in the bosonic case
$$H^0\phi (\ux )=2\phi (x_1+1,x_2)+2\phi (x_1,x_2-1),\;
\ux\in\partial\z2_{\geq},$$
while in the fermionic case $H$ is considered on functions
$f$ vanishing on $\partial\Z^2_{\geq}$.
\pmn

\noindent{\bf Remark.} The method used in this paper had been specifically
designed for bosonic and fermionic systems. An extension of our results to the
Maxwell--Boltzmann statistics is possible but would require additional technical
constructions.
\pmn

The interaction potential $U:\;\ux\in\Z^2_{\geq}\mapsto
\R$ is a fixed real-valued function vanishing when $x_1-x_2$ exceeds a given value
$d<\infty$:
$$U(\ux)=0,\;\hbox{ if }\;x_1-x_2>d.\eqno(1.4)$$

In addition, there is given a family of real IID random variables
$V(x,\;\cdot\;)$, $x\in\Z$,
representing the external field. Constant $g$ (the amplitude parameter) will be assumed
big, but may be of positive or negative sign.

As a working approximation for $H$ we consider a Hermitian
$|\Lam |\times |\Lam |$ matrix $H_\Lam\left(
=H^{(2)}_{\Lam,U,V,g}(\omega )\right)$ where $\Lam\subset\Z^2_{\geq}$
is a finite set of cardinality $|\Lam |$. Matrix $H_\Lam$ is of the form
$H^0_\Lam+U+g(V_1+V_2)$ and respresents an LSO in $\Lam$:
$$
\begin{array}{l}H_\Lam\phi (\ux)=H^0_\Lam\phi (\ux)+
\left[\big(U+gV_1+gV_2\big)\phi\right](\ux )\\
\;\;=\Big[\phi (x_1-1,x_2){\mathbf 1}_\Lam (x_1-1,x_2)
+\phi (x_1+1,x_2){\mathbf 1}_\Lam (x_1+1,x_2)\\
\;\;\;\;+\phi (x_1,x_2-1){\mathbf 1}_\Lam (x_1,x_2-1)
+\phi (x_1,x_2+1){\mathbf 1}_\Lam (x_1,x_2+1)\Big]\\
\;\;+\left[U(\ux )+gV(x_1,\omega )+gV(x_2,\omega )\right]
\phi (\ux ),\;\;\ux=(x_1,x_2)\in\Lam,\end{array}\eqno(1.5)$$
${\mathbf 1}_\Lam$ being the indicator function of $\Lam$. In fact,
we focus on lattice squares or their intersections with $\Z^2_{\geq}$, and use the
notation
$$\Lam_L(\uu )=\Big([u_1-L,u_1+L]
\times [u_2-L',u_2+L']\Big)\cap\Z^2_{\geq}.\eqno(1.6)$$
Such a set is called a (lattice) sub-square.

Given a finite set $\Lam^{(1)}\subset\Z$, we can also consider a single-particle LSO
$H^{(1)}_{\Lam^{(1)}}\left(= H^{(1)}_{\Lam^{(1)},V,g}\right)$ of the form
$$\begin{array}{r}
H^{(1)}_{\Lam^{(1)}}\phi (x)=\Big[\phi (x-1){\mathbf 1}_{\Lam^{(1)}}(x)
+\phi (x+1){\mathbf 1}_{\Lam^{(1)}}(x+1)\Big]\qquad{}\\
+gV(x,\omega )\phi (x),\;x\in\Lam^{(1)}.\end{array}\eqno(1.7)$$
Of particular interest to us are (lattice) segments:
$$\Lam^{(1)}=\Lam^{(1)}_L(u)=[u-L,u+L]\cap\Z.\eqno (1.8)$$
Matrix $H^{(1)}_{\Lam^{(1)}_L(u)}$ gives a finite-volume approximation to a
single-particle LSO $H^{(1)}$ on $\Z$:
$$H^{(1)}\phi (x)=\Big[\phi (x-1)
+\phi (x+1)\Big]+gV(x,\omega )\phi (x),\;x\in\Z,\eqno(1.9)$$
which acts in the single-particle Hilbert space $\ell_2(\Z )$.

Next, a system of two particles in a finite volume $\Lam\in\Z^2_{\geq}$ with no
interaction is described by the LSO $H^{\rm{n-i}}_\Lam\left(
=H^{(2),{\rm{n-i}}}_{\Lam,U,V,g}\right)$ of the form $H^0_\Lam+g(V_1+V_2)$:
$$\begin{array}{l}
H^{\rm{n-i}}_\Lam\phi (\ux)=H^0_\Lam\phi (\ux)+ g\left[(V_1+V_2\big)\phi\right](\ux )\\
\;\;=\Big[\phi (x_1-1,x_2){\mathbf 1}_\Lam (x_1-1,x_2)
+\phi (x_1+1,x_2){\mathbf 1}_\Lam (x_1+1,x_2)\\
\qquad +\phi (x_1,x_2-1){\mathbf 1}_\Lam (x_1,x_2-1)
+\phi (x_1,x_2+1){\mathbf 1}_\Lam (x_1,x_2+1)\Big]\\
\qquad +g\left[V(x_1,\omega )+V(x_2,\omega )\right]
\phi (\ux ),\;\;\ux=(x_1,x_2)\in\Lam .\end{array}\eqno (1.10)$$
In this paper we work with matrices $H^{\rm{n-i}}_\Lam$ where
$\Lam$ is a (lattice) square
$\Lam_L(\uu )=\Lam^{(1)}_L(u_1)\times\Lam^{(1)}_L(u_2)$ lying inside
$\z2_{\geq}$, where segments $\Lam^{(1)}_L(u_j)$ are as in (1..).
In this case we can use the straightforward representation
$$ H^{\rm{n-i}}_{\Lam_L(\uu )}=H^{(1)}_{\Lam^{(1)}_L(u_1)}
\otimes I_{\Lam^{(1)}_L(u_2)}
+ I_{\Lam^{(1)}_L(u_1)}\otimes H^{(1)}_{\Lam^{(1)}_L(u_2)}.
\eqno(1.11)$$
Of course, the spectrum of matrix $H^{\rm{n-i}}_{\Lam_L(\uu )}$ will be formed by the
sums of the eigen-values (EVs) of
$H^{(1)}_{\Lam^{(1)}_L(u_1)}$ and $H^{(1)}_{\Lam^{(1)}_L(u_2)}$.

This brings us to the observation that the principal difference between a single-particle
random LSO (1.1) on $\Z^2$ and a two-particle LSO (1.2) on
$\Z^2_{\geq}$ is that the values of the external potential field
$$\ux\mapsto g\big[V_1(x_1,\omega ) +V_2(x_2,\omega )\big]\eqno (1.12)$$
in (1.2) are `strongly' dependent. For example, for any two points $\ux =(x_1,x_2)$ and
$\ux' =(x_1+a,x_2)$ from $\Z^2_{\geq}$, with $a \geq 1$, the values
$$gV(x_1,\omega)+gV(x_2,\omega)\;\text{ and } gV(x_1+a,\omega)+
gV(x_2,\omega)$$ are coupled, as RV $V(x_2,\om)$ is present in both sums. On the other
hand, LSOs (1.1) and (1.2) bear essential similarities, owing to the fact that the
approximating matrix $H_\Lam$, for a square $\Lam_L(\uu )$ `deeply inside'
$\z2_{\geq}$, coincides with $ H^{\rm{n-i}}_{\Lam_L(\uu )}$. This allows
us to apply a number of results and techniques from the single-particle MSA scheme, while
some other key points of the scheme have to be modified or extended.

\subsection{The main result}
\label{ssec:1.2}

Our assumptions throughout the paper are as follows.
\pmn

(A) {\sl RV's $V(x,\;\cdot\;)$, $x\in\Z$, are IID and have a probability
density function (PDF) $f$ which is bounded on $\R$:
$$||f||_{\infty}=\sup\;[f(y):\;y\in\R]<\infty,$$
and is such that the characteristic function
$$\esm{ e^{itV(x,\;\cdot\;)}}=\int_{\R}{\rm d}y\,e^{ity}f(y)$$
admits the bound
$$\left|\esm{  e^{itV(x,\;\cdot\;)} } \right|\leq
be^{-a|t|},\eqno (1.13)$$ where $a>0$ and $b\geq 1$ are constants.}

(B) $U$ {\sl is a real bounded function on $\Z^2_\geq $
satisfying} (1.4).
\pmn

Bound (1.13) implies that PDF $f(y)$, $y\in\R$, admits the analytic continuation into a
strip $\left\{z\in\C:\;
\left|{\rm{Im}}\;z\right|<a\right\}$.

As was indicated, the statistics of the system is defined by the type of the boundary
conditions on $\partial\Z^2_{\geq}$. In both cases, LSO $H$ formally defined by (1.2) is
initially considered on the set of functions $f$ with compact support. Here, with
probability one, it is essentially self-adjoint, and we take its self-adjoint extension
which is again denoted by $H(=H^{(2)}_{U,V,g}(\om ))$. Theorem 1.1 below addresses both
cases.
\psn
\vskip 1 truecm

{\bf Theorem 1.1} $\quad$ {\sl Assume that conditions} (A)
{\sl and} (B) {\sl are fulfilled.
Then there exists $g_0\in (0,\infty )$ such that if
$|g|\geq g_0$ then LSO $H$ in} (1.2)
{\sl satisfies the following property. With probability one,}

(a) {\sl the spectrum of $H$ is pure point:
$\sigma (H)$ $=$ $\sigma_{\rm{pp}}(H)$, and}

(b) $\forall$ {\sl eigen-value}
$E\in\sigma_{\rm{pp}}(H)$, {\sl every corresponding EF
$\psi (\ux;E)$ exhibits an exponential decay:}
$$\limsup\limits_{||\ux ||\to\infty}
\frac{\log\;|\psi (\ux;E)|}{||\ux ||}=-m<0.\eqno (1.14)
$$
{\sl Here,
$||\ux ||$ stands for the Euclidean norm
$\big(|x_1|^2+|x_2|^2\big)^{1/2}$; the value $m$
($=m(\psi (\;\cdot\;;E))$) is called the mass
(of eigen-function $\psi (\;\cdot\;;E)$).}
\psn
\vskip 1 truecm

The threshold $g_0$ in Theorem 1.1 can be assessed in terms of the sup--norm
$|f|_\infty$, the constants $a$ and $b$ in Assumption (A) and the radius of interaction
$d$ and the maximum $\max\;\big[|U (r)|:\;r\in\Z_+\big]$ in Assumption (B).

Throughout the paper, symbol
$\qed$ is used to mark the end of a proof.

\section{Wegner-type estimates}

One of the key ingredients of MSA is an estimate of the probability to find an EV of LSO
$H_\Lam$ (see (1.5)) in an interval $(E_0-r,E_0+r )$. The Wegner
estimate, used for IID values of the external potential, does not apply directly to our
problem. So, we need an analog of the Wegner estimate of the density of states. For
definiteness, we assume that $\Lam$ is a lattice rectangle.

Let
${\rm d}\kappa_{\Lam}(\lam)$ be the averaged spectral measure of
$H_\Lam $ such that
$$\esm{ \left\langle\delta_\uu,\varphi\left(H_\Lam
\right)\delta_\uu\right\rangle }=\int\varphi (\lam)
{\rm d}\kappa_{\Lam}(\lam),\;\;\uu\in\Lam,\eqno (2.1)$$
for any bounded test function $\varphi$. Here and below,
$\delta_\uu$ stands for the Dirac's delta, and
$\langle\;\cdot\;,\;\cdot\;\rangle$ and $||\;\cdot\;||$ denote
the inner product and the norm in $\ell^2(\Z^2)$. It is well-known that measure
${\rm d}\kappa_{\Lam}$ is independent of the choice
of an element
$\phi =\sum\limits_{\uu\in\Lam}\left\langle\phi,\delta_{\uu}
\right\rangle\delta_{\uu}$ with $||\phi ||=1$: for any such $\phi$,
$$
\esm{ \left\langle\phi,\varphi\left(H_\Lam  \right)\phi\right\rangle }
= \int\varphi (\lam) {\rm d}\kappa_{\Lam}(\lam).
\eqno(2.2)
$$
Actually, ${\rm d}\kappa_\Lam$  is a normalised (i.e. a probability) measure on $\R$. Let
${\wh k}_{\Lam}(t)$ be its inverse Fourier transform
(the characteristic function, in a probabilistic terminology),
$${\wh k}_{\Lam}(t)=\esm{ \langle\delta_{\uu},
e^{itH_\Lam }\delta_{\uu}\rangle },\;\;\uu\in
\Lam,\;t\in\R.\eqno(2.3)$$
then
$$k_{\Lam}(s)=\frac{1}{2\pi}\int_{\R} e^{-ist}
{\wh k}_{\Lam}(t){\rm d}t.\eqno(2.4)$$
\pmn

\begin{Lem}\label{DecayFourier} The quantity
${\wh k}_{\uu,\Lam}$ defined in {\rm{(2.3)}} obeys
$$\left|{\wh k}_{\uu,\Lam}(t)\right|
\leq e^{-B|t|},\;\;\uu\in\Lam,\;t\in\R,\eqno(2.5)$$
where
$$B=2(a|g|-b-1),\eqno(2.6)$$
independently of $\Lam$. Therefore, $k_\Lam(\lam)$ is analytic in a strip around real
line, so it has a bounded derivative on any interval.
\end{Lem}
\pmn
{\bf Proof of Lemma \ref{DecayFourier}.} For definiteness, we focus
on the fermionic case. To estimate $\left|{\wh k}_{\uu,\Lam}(t)\right|$, we use
Molchanov's formula expressing matrix elements of
$e^{itH_\Lam }$ in terms of the integral over trajectories
of a Markov jump process on the time interval $\big[\,0,\,|t|\,\big]$. Namely,
$$\begin{array}{l}
\langle\delta_{\uu},e^{itH_\Lam }\delta_{\uu}\rangle\\
= e^{4|t|} \bfE_\uu\left({\mathbf 1}(\uX (t)=\uu)
i^{K(|t|)({\rm{sign}}\;t)} \exp\;\left[i({\rm{sign}}\;t)
\diy{\int\limits_0^{|t|}} W(\uX (s)){\rm d}s\right]\right).
\end{array}$$
Here $\{\uX (s), s\geq 0\}$ is the continuous-time Markov jump process on $\Lam$, with
holding times of rate $2$, equiprobable jumps to four nearest neighbour sites and
Dirichlet's boundary conditions outside $\Lam$. Next, $\bfE_\uu$ denotes the expectation
generated by the distribution of the process when the initial site is $\uu$. Further,
$K(|t|)$ ($=K\big(|t|;\{\uX (s)\}\big)$ is the number of
jumps of $\{\uX (s)\}$ between times 0 and $|t|$. Now,
$$\begin{array}{l}
{\wh k}_{\uu,\Lam}(t)\\
=e^{4|t|}\esm{ \esms{\uu}{ {\mathbf 1}(\uX (t) =\uu)i^{K(|t|)
({\rm{sign}}\;t)}\exp\;\left(i({\rm{sign}}\;t)
\diy{ \int\limits_0^{|t|}} W(\uX (s)){\rm d}s  \right)} } \\

= e^{4|t|}\esms{\uu}
{
\esm{ {\mathbf 1}(\uX (t) =\uu)i^{K(|t|) ({\rm{sign}}\;t)}
\exp\;\left(i({\rm{sign}}\;t)
\diy{ \int\limits_0^{|t|}} W(\uX (s)){\rm d}s  \right)}}\end{array}$$

the change of order of integration is justified by the boundedness of the integrand.

For simplicity we assume from now on that $t>0$. In our case,
$$W(\uu) = V(u_1)+ V(u_2) + U(\uu),\;
\;\uu=(u_1,u_2).$$
Given trajectory $\uX (s)$, $s\geq 0$, the values $K(t)$ and $U(\uu)$, $\uu\in\Lam$, are
non-random. Hence, the internal expectation
$$\begin{array}{l}\E\left({\mathbf 1}(\uX (t)=\uu)i^{K(t)}
\exp\;\left[i\diy{\int_0^t}W(\uX (s)){\rm d}s\right]\right)\\
           \;\;{}={\mathbf 1}(\uX (t)=\uu)i^{K(t)}
\exp\left[i\diy{\int_0^t}U(\uX (s)){\rm d}s\right]\E
\exp\left(ig\sum\limits_{j=1}^2\diy{\int_0^t}V(X_j(s))
{\rm d}s\right),\end{array}$$
where $X_1(s)$, $X_2(s)$ are the components of $\uX(s)$.

Write
$$\sum_{j=1}^2\int_0^t  V(X_j(s)){\rm d}s
=\sum\limits_{z\in\Z}V(z)\sum_{j=1}^2 \tau^j(z),$$
where $\tau^j(z)$ is the time spent at $z$ by process $\{X_j(s)\}$ between 0 and $t $.
This yields
$$\begin{array}{l}\E\left[\bfE_\uu\left({\mathbf 1}(\uX (t)=\uu)i^{K(t )}
\exp\;\left[i\diy{\int_0^t}W(\uX (s)){\rm d}s\right]\right)\right]\\
\quad{}=\bfE_\uu\left[{\mathbf 1}(\uX (t)=\uu )i^{K(t )}
\E\;\exp\;\left(ig\sum\limits_{z\in\Z}V(z)\sum\limits_{j=1}^2
\tau_j(z)\right)\right].\end{array}$$
Then
$$\begin{array}{l}\left|\;\E\left[
\bfE_\uu\left({\mathbf 1}(\uX (t)=\uu)i^{K(t )}
\exp\;\left[i\diy{\int_0^t}  W(\uX (s)){\rm d}s\right]\;\right)
\right]\;\right|\\
\quad{}\leq
\bfE_\uu\;\left[{\mathbf 1}(\uX (t)=\uu)
\left|\E\;\exp\;\left(ig \sum\limits_{z\in\Z}V(z)\sum\limits_{j=1}^2
\tau^j(z)\right)\right|\;\right]\\
\quad{}=\bfE_\uu\left[\;{\mathbf 1}(\uX (t)=\uu )
\prod\limits_{z\in\Z}\left|
\E e^{igV(z)(\tau^1(z) + \tau^2(z))}\right|\;\right];\end{array}$$
the last equality holds as RVs $V(z)$ are independent for different $z$.

By (1.13), the last expression is
$$\leq\bfE_\uu\left[{\mathbf 1}(\uX (t)=\uu )
b^{M(t )}\exp\;\left(-a|g|\sum_{z\in\Z}
\left(\tau^1(z) + \tau^2(z)\right)\right)\right]
$$
which equals $e^{-a|g|\cdot 2t }\bfE_\uu b^{M(t )}$, as the sum $\sum\limits_{z\in\Z}
\left(\tau^1(z) + \tau^2(z)\right)=2t $.
Here $M(t )$ ($=M\big(t ;\{\uX (s)\}\big)$ is the total number of sites in $\Z$ visited
by processes $\{X_j(s)\}$, $j=1,2$, between times 0 and $t $. Since $M(t )\leq K(t )$, we
have that
$$e^{-2at}\bfE_\uu b^{M(t )}\leq e^{-2a|g|t}
\bfE_\uu b^{K(t )} = e^{-2t (a|g|-b+1)}.$$

For the matrix elements $\langle\delta_{\uu}, e^{itH_\Lam }\delta_{\uu}\rangle$ we get
the bound
$$
\left|\langle\delta_{\uu},
e^{itH_\Lam }\delta_{\uu}\rangle\right|\leq e^{-2t (a|g|-b-1)}.
$$
This completes the proof of Lemma  \ref{DecayFourier}. $\qed$
\pmn

{\bf Remark.} Molchanov's formula
has been used in \cite{CL}, Proposition VI.3.1, to prove analyticity of the integrated
density of states in the single-particle Anderson model with an IID random potential of
the same type as in the present paper. As we will see, path integration techniques can be
adapted to multi-particle lattice systems in any dimension.
\pmn

We see that ${\rm d}\kappa_\Lam(\lam)$ admits a density:
${\rm d}\kappa_\Lam(\lam)=k_{\Lam}(\lambda){\rm d}\lambda$.
\pmn

{\bf Theorem 2.1. \label{WegnerType}(A Wegner-type estimate)}
{\sl Consider LSO $H_\Lam$, as in {\rm{(1.5)}},
with $\Lam =\Lam_{L_1,L_2}(\ux )=\Lam_{L_1}(x_1)\times
\Lam_{L_2}(x_2)$. Under
conditions {\rm{(A)}} and {\rm{(B)}},
$\forall$ $E\in\R$, $L_1,L_2\geq 1$, $r>0$, and $\forall$
$\ux=(x_1,x_2)\in\Z^2$, probability $\P\;\Big\{{\rm{dist}}
\left[\;E,\sigma(H_{\Lam}) \right]<r\Big\}$ satisfies
$$\P\;\Big\{{\rm{dist}}\left[\;E,\sigma(H_{\Lam}) \right]
<r\Big\}\leq \frac{2}{\pi B}(2L_1+1)(2L_2+1)r,$$ where $B$ is the same as in Equation
{\rm{(2.6)}}. In particular, for $r =e^{-(L_1\wedge L_2)^{\beta/2}}$,
$$\P\;\Big\{{\rm{dist}}\left[\;E,\sigma(H_{\Lam}) \right]
< e^{-(L_1L_2)^{\beta/2}}\Big\}
\leq \frac{2}{\pi B}(2L_1+1)(2L_2+1)e^{-(L_1\wedge L_2)^{\beta/2}}.
\eqno (2.7)$$
Here and below, $L_1\wedge L_2 = \min\{L_1,L_2\}$.}
\pmn
{\bf Proof of Theorem 2.1.} We begin with an elementary
inequality (cf. \cite{CL}). Let
$\Pi_{(E-r,E+r)}^{\Lam}$ be the spectral projection
on the the subspace spanned by the corresponding EFs of $H_\Lam $. Then
$$\P\Big\{{\rm{dist}}\;\left[E,\sigma\left(H_\Lam \right)\right]<r
\Big\}\leq \E\;{\rm{tr}}\;\Pi_{(E- r,E+ r)}^{\Lam}.\eqno (2.8)$$

Further, in the Dirac's delta-basis:
$${\rm{tr}}\;\Pi_{(E- r,E+ r)}^{\Lam}
=\sum_{\uu\in\Lam}
\left\langle\delta_\uu,
\Pi_{(E- r,E+ r)}^{\Lam}\delta_\uu\right\rangle ,$$
and
$$\E\;{\rm{tr}}\;\Pi_{(E- r,E+ r)}^{\Lam}
=\sum\limits_{\uu\in\Lam}
\int_{E- r}^{E+ r}
k_{\Lam}(s){\rm d}s.$$ The assertion of Theorem 2.1 now follows easily from Lemma
\ref{DecayFourier} and Equations (2.3)--(2.4).
$\qquad\qed$
\pmn

We will also need a variant of the Wegner-type estimate where either the horizontal or
vertical projection sample of the potential is fixed. In Lemma \ref{DecayFourierCond} and
Theorem 2.2 it is assumed that the lattice rectangle $\Lam =\Lam^{(1)}_1 \times
\Lam^{(1)}_2$ has $\Lam^{(1)}_1\cap\Lam^{(1)}_2=\emptyset$.
In Lemma \ref{DecayFourierCond} we consider the conditional expectation
${\wh k}_{\uu,\Lam}\big(t\big|{\mathfrak V}(\Lam^{(1)}_2)\big)$:
$${\wh k}_{\Lam}\big(t\big|{\mathfrak V}(\Lam^{(1)}_2)\big)
=\E\left[\langle\delta_{\uu},
e^{itH_\Lam }\delta_{\uu}\rangle \,|\,{\mathfrak V}(\Lam^{(1)}_2)\right],
\;\;\uu\in\Lam,\;t\in\R,\eqno(2.9)$$
where the sigma-algebra
${\mathfrak V}(\Lam^{(1)}_2)=\{V(x,\cdot),\;x\in\Lam^{(1)}_2\}$ is
generated by the values of the potential potential over segment
$\Lam^{(1)}_2$.
\pmn

\begin{Lem}\label{DecayFourierCond} The quantity
${\wh k}_{\uu,\Lam}\big(t\big|{\mathfrak V}(\Lam^{(1)}_2)\big)$
defined in {\rm{(2.9)}} obeys
$$\sup \left|{\wh k}_{\uu,\Lam}
\big(t\big|{\mathfrak V}(\Lam^{(1)}_2)\big)\right|\leq e^{-Bt/2},\;\;
\uu\in\Lam,\;t\in\R,\eqno(2.10)$$
independently of $\Lam$. Here, as in {\rm{(2.6)}}, $B=2(a|g|-b-1)$. Therefore,
$k_\Lam(\lam)$ is analytic in a strip
around real line, so it has a bounded derivative on any interval.
\end{Lem}
\pmn

Lemma 2.2 is proved in the same way as Lemma 2.1. A direct corollary of Lemma 2.2 is
\pmn

{\bf Theorem 2.2. \label{WegnerTypeCond}(A conditional Wegner-type
estimate)} {\sl For LSO $H_\Lam$, as in {\rm{(1.5)}}, with $\Lam =\Lam_{L_1,L_2}(\ux
)=\Lam_{L_1}(x_1)\times\Lam_{L_2}(x_2)$ and
$\Lam_{L_1}(x_1)\cap\Lam_{L_2}(x_2)=\emptyset$, under assumptions {\rm{(A)}} and
{\rm{(B)}},
$\forall$ $E\in\R$, $L_1,L_2\geq 1$, $r>0$, and $\forall$
$\ux =(x_1,x_2)\in\Z^2$,
the conditional probability $\P\;\Big\{{\rm{dist}}
\left[\;E,\sigma(H_{\Lam})\right]
<r\big|{\mathfrak V}(\Lam^{(1)}_2)\Big\}$ satisfies
$$\sup \;\P\;\Big\{{\rm{dist}}\left[\;E,\sigma(H_{\Lam})\right]
<r\big|{\mathfrak V}(\Lam^{(1)}_2)\Big\}\leq
\frac{4}{\pi B}(2L_1+1) (2L_2+1)r,$$
where $B$ is the same as in Equation {\rm{(2.6)}}. In particular, for $r=e^{-(L_1\wedge
L_2)^{\beta/2}}$,
$$\begin{array}{r}
\sup \;\P\Big\{{\rm{dist}}\left[\;E, \sigma(H_{\Lam})\right]
< e^{-(L_1L_2)^{\beta/2}}\big|{\mathfrak V}(\Lam^{(1)}_2)\Big\}
\qquad{}\qquad{}
\\
\leq\diy{\frac{4}{\pi B}}(2L_1+1)(2L_2+1)
e^{-(L_1\wedge L_2)^{\beta/2}}.\end{array}\eqno (2.11)$$}
\pmn

{\bf Remark.} Obviously, similar estimate holds for the
conditional expectation with respect to the sigma-algebra
${\mathfrak V}(\Lam^{(1)}_1)=\{V(x,\cdot), x\in\Lam^{(1)}_1\}$.
\pmn

We conclude this section with the statement which is a straightforward refinement of
Theorem 2.2 and can be proved in a similar fashion.
\pmn

{\bf Theorem 2.3.} \label{WegnerTypeCond}
{\sl Consider segments $I_1=[a_1, a_1+L'_1]\cap \Z$,
$I_2=[a_2, a_2+L''_1]\cap \Z$, $J_1=[c_1, c_1+L_2']\cap \Z$ and
$J_2=[c_2, c_2+L_2'']\cap \Z$. Assume that
$$
\text{ either } \,\, I_1 \cap (J_1 \cup I_2 \cup J_2) = \emptyset
\text{ or } J_1 \cap( I_1 \cup I_2 \cup J_2 ) = \emptyset.
$$
Set $\Lam' = I_1\times J_1$, $\Lam'' = I_2\times J_2$. Let
${\mathfrak V}(I_2\cup J_2)$ stand for the sigma-algebra
$\{V(x), x\in I_2\cup J_2\}$. Consider an arbitrary function
${\mathcal E}$ measurable relative to ${\mathfrak V}(I_2\cup J_2)$.
Then
$$\sup \;\P\;\Big\{{\rm{dist}}\left[\;
{\mathcal E},\sigma(H_{\Lam'})\right]
<r\big|{\mathfrak V}(I_2\cup J_2)\Big\}\leq \frac{4}{\pi B}(2L'_1+1)
(2L'_2+1)r.$$
In particular, for $r=e^{-(L_1'\wedge L'_2)^{\beta/2}}$,
$$\begin{array}{r}
\sup \;\P\;\Big\{{\rm{dist}}\left[\;{\mathcal E},\sigma(H_{\Lam'})\right]
<r\big|{\mathfrak V}(I_2\cup J_2)\Big\}\qquad{} \qquad{} \qquad{} \\
\leq \diy{\frac{4}{\pi B}}(2L'_1+1) (2L'_2+1)
e^{-(L_1'\wedge L'_2)^{\beta/2}}.\end{array}
\eqno(2.12)$$}

\begin{Lem}\label{Lemma 2.3}
Let $\Lam'=\Lam'_{L}=I_1 \times J_1$,
$\Lam''=\Lam''_{L}=I_2 \times J_2$, be two sub-squares with
$I_j=[a_j,b_j]$, $J_j=[c_j,d_j]$, $j=1,2$ and such that
$$
\Lam' \cap \cD_d \neq \emptyset, \; \Lam'' \cap \cD_d \neq \emptyset.
$$
Assume that the max-norm distance
$$
d_\infty (\Lam',\Lam'' ) > 5L
$$
and that $L>d$. Then the coordinate projections of $\Lam'$ are disjoint from those of
$\Lam''$: $\big(I_1\cup J_1\big)\cap
\big(I_2\cup J_2)=\emptyset$, and so the potential samples in $\Lam'$
and $\Lam''$ are independent.
\end{Lem}
\pn
{\bf Proof of Lemma \ref{Lemma 2.3}.}
Indeed, since $\Lam' \cap \cD_d \neq \emptyset$, then $(a_2,d_2)\in \cD_d$, so that $a_2
- d_2 \leq d$. Further, $d_2 - c_2 \leq 2L$, so we have
$$
c_2 \geq d_2 - 2L \geq a_2 - d - 2L.
$$
On the other hand, since $\Lam' \cap \Z^2_{\geq}$, we have $(b_1,c_1)\in\Z^2_\geq$, so
that $c_1 \leq b_1$. Therefore,
$$
d_1 \leq c_1 + 2L \leq b_1 + 2L.
$$
Combining the above inequalities, we see that
$$
c_2 - d_1 \geq (a_2 - d - 2L)  - (b_1 - 2L)
= (a_2 - b_1) - 4L - d > 5L - (4L+d)>0,
$$
so that $J_1 \cap J_2 = \emptyset$. Taking into account that $I_1\cap I_2=\emptyset$, we
conclude that
$$
(I_1 \cup J_1) \cap (I_2 \cup J_2) = \emptyset.
\qed
$$

\begin{Def} We call a pair of  sub-squares $\Lam'$,
$\Lam''$  $L$-distant ($L${\rm -D},
for short), if
$$
\dinf(\Lam', \Lam'') > 8L.
$$
\end{Def}


\begin{Lem}\label{Lemma 2.4}
Let $\Lam' = (I_1 \times J_1) \cap \Z_\geq^2$ and
$\Lam'' = (I_2 \times J_2) \cap \Z^2_{\geq}$
be two  sub-squares in $\Z_\geq^2$. Assume that:
{\rm(a)} the lengths of four segments $I_j$ and $J_j$
is $\leq 2L$, $j=1,2$, and
{\rm(b)} $\Lam', \Lam''$ are $L$-D, i.e. $d_\infty(\Lam',\Lam'') > 8 L$.

Then either

{\rm(A)} at least one of sub-squares $\Lam', \Lam''$ is off-diagonal
(and hence is a square), in which case
at least one of their coordinate projections is disjoint from the three others,
\psn
or

{\rm(B)} the projections of $\Lam'$ are disjoint from those
of $\Lam''$: $(I_1 \cup J_1) \cap (I_2 \cup J_2) = \emptyset$.
\end{Lem}
\pn
{\bf Proof of Lemma \ref{Lemma 2.4}.} Denote by $\cJ$ the union
of four segements
$I_1 \cup J_1 \cup I_2 \cup J_2$ and call it disconnected
if (i) there exists a segment, among the four, disjoint from the three others, or (ii)
there are two pairs of segments disjoint from each other, although within each pair the
segments have non-empty intersections. Otherwise, $\cJ$ is called connected.

First, note that had set $\cJ$ been connected, its diameter would have been bounded by
$8L$, since each interval has length $\leq 2L$. Then we would have had
$$
\dinf(I_1, I_2) \leq 4L, \dinf (J_1, J_2)
\leq 8L \Rightarrow \dinf(\Lam', \Lam'') \leq 8L,
$$
which is impossible by assumption (b).

Thus, assume that $\cJ$ is disconnected. It is straightforwad that in case (i) the
assertion (A) of the Lemma 2.4 holds true. Hence we only have to show that in case (ii),
both $\Lam'$ and $\Lam''$ are diagonal sub-squares.

In case (ii) we call the unions of segments within a given pair a connected component (of
$\cJ$). By assumption (b), either $\dist(I_1,I_2) > 8L$ or $\dist(J_1,J_2) > 8L$. For
definiteness, suppose that $\dist(I_1,I_2) > 8L$. Then $I_1$ is disjoint from $I_2$, and
the connected component of $\cJ$ containing $I_1$ should include either $J_1$ or $J_2$.
Suppose first that
$$I_1 \cap J_1\neq \emptyset,\;\hbox{ and }\;I_2 \cap J_2\neq \emptyset,
\eqno(2.13)$$
then $\Lam' \cap \cD_d \neq \emptyset$ and
$\Lam'' \cap \cD_d \neq \emptyset$. By virtue of property (b),
Lemma 2.3 applies, and assertion (B) in this case holds true.

Now suppose that
$$
I_1 \cap J_2 \neq \emptyset\;\hbox{ and }\;I_2 \cap J_1 \neq \emptyset.
$$
Then
$$I_1 \cap J_1 = \emptyset,\;I_2 \cap J_2
= \emptyset\;\hbox{ and }\;J_1 \cap J_2 = \emptyset.\eqno(2.14)$$
We see that both $\Lam'$ and $\Lam''$ are off-diagonal squares. Write
$I_j=[a_j,b_j]$, $J_j=[c_j,d_j]$, $j=1,2$. Since $I_2\cap I_1=\emptyset$,
we can assume without loss of generality that
$$a_1 < b_1 < a_2 < b_2.$$
Further, $(a_1,d_1)\in \Lam' \subset \Z^2_>$, so that
$$c_1<d_1<a_2<b_2\;\hbox{ implying that}\;I_2\cap J_1=\emptyset.
$$
In turn, this yields
$$I_2 \cap (I_1 \cup J_1) = \emptyset.$$
But as $I_2 \cap J_2 = \emptyset$ (see (2.14)), then
$$
I_2 \cap (J_2 \cup I_1 \cup J_1) = \emptyset,
$$
which is impossible: we are in case (ii), so no interval among $I_1$, $I_2$, $J_1$, $J_2$
is disjoint from the remaining three. This completes the proof of Lemma 2.4.
\noindent\qed

\section{The MSA scheme: a single-particle case}
\pmn

Throughout this section we assume that condition (A) holds, although the scheme works for
a much larger class of IID RVs $V(x,\;\cdot\;)$, $x\in \Z$.
(In fact, the MSA scheme does not even require dimension one.)

For reader's convenience, we reproduce here the principal points of the proof of
localisation given in \cite{vDK1}. To simplify the future adaptation of the MSA scheme to
the case of two particles, we choose particular values of parameters $p$, $q$, $\alpha$
and $\beta$ figuring in the specification of the scheme. This does not reduce the
generality of the construction.

\begin{Def} Fix $\beta=\vbeta$. Given $E\in\R$,
a segment $\Lam^{(1)}_L(x)=[x-L,x+L]\cap\Z$,
$x\in\Z$, is called $E$-resonant
($E${\rm-R}, for short) if the spectrum
$\sigma(H^{(1)}_{\Lam^{(1)}_L(x)})$
of $H^{(1)}_{\Lam^{(1)}_L(x)}$, the single--particle LSO in $\Lam^{(1)}_L(x)$
(see {\rm{(1.7)}}), satisfies
$$
{\rm dist}\left[E, \sigma(H^{(1)}_{\Lam^{(1)}_L(x)})\right]
< e^{-L^\beta}.\eqno (3.1)$$
Otherwise, $\Lam^{(1)}_L(x )$ is called $E$-non-resonant
($E$-{\rm{NR}}).
\end{Def}

\begin{Def} Given $E\in\R$ and $m>0$, a segment
$\Lam^{(1)}_L(x ) = [x-L,x+L] \cap \Z$, $x\in\Z$, is called
$(E,m)$-non-singular ($(E,m)${\rm -NS}, for short) if
$$\max_{u:\,|u-x|=L}\left|G^{(1)}_{\Lam^{(1)}_L(x)}(x,u; E)\right|
\leq e^{-mL}. \eqno (3.2)
$$
Otherwise it is called $(E,m)$-singular ($(E,m)$-{\rm S}). Here,
$G^{(1)}_{\Lam^{(1)}_L(x )}(y ,u ; E)$, $y,u\in \Lam^{(1)}_L(x)$,
stands for the Green's function of $H^{(1)}_{\Lam^{(1)}_L(x)}$:
$$
G^{(1)}_{\Lam^{(1)}_L(x )}(y ,u ; E) = \left\langle \left( H^{(1)}_{\Lam^{(1)}_L(x)} - E
\right)^{-1} \delta_y, \delta_u
\right\rangle, \;\;y,u\in \Lam^{(1)}_L(x).\eqno (3.3)$$
\end{Def}
\pmn

In Theorems 3.1 and 3.2 we consider intervals $I\subset \R$ of length $\leq 1$. However,
the statements of both theorems can be easily extended to any finite interval.
\pmn

\begin{Thm}\label{vDK22} Let $I\subset\R$ be an
interval of length $\leq 1$. Given
$L_0>0$, $m_0>0$, $p=\vp$, $q=\vq$
and $\beta=\vbeta$, consider the following properties
{\bf{(S1.0)}} and {\bf{(S2.0)}} of single-particle LSOs
$H^{(1)}_{\Lam^{(1)}_L}$ in {\rm{(1.7)}}:
\psn
$$\begin{array}{l}\hbox{{\bf (S1.0)} $\quad$
$\forall$ $x,y\in\Z$ and disjoint segments $\Lam^{(1)}_{L_0}(x)$
and $\Lam^{(1)}_{L_0}(y)$,}\\
\P\;\Big\{\forall \, E\in I:\;{\rm{both}}\;\Lam^{(1)}_{L_0}(x )
\;{\rm{and}}\;\Lam^{(1)}_{L_0}(y )
\;{\rm{are}}\;(E,m_0){\rm -S}\Big\}<L_0^{-2p}.
\end{array}\eqno (3.4)$$
\psn
$$
\begin{array}{l}\hbox{{\bf (S2.1)} $\quad$ $\forall$ $L\geq L_0$
and $\forall$ $E$ with
${\rm dist}\left[E,I\right]\leq\half e^{-L^\beta}$,}\\
\P\;\Big\{
 \; \Lam^{(1)}_{L}(x) \text{ is } E\text{\rm-R}
\Big\}< L^{-q}.\end{array}\qquad\quad\;\;\;\,\eqno (3.5)$$
\psn
Take $\alpha = \valpha$. Next set
$L_{k+1}=L_k^\alpha$, $k=0, 1, \dots$. Given a number
$m\in(0,m_0)$, $\exists$ $Q^0=Q^0(m_0,m) <\infty$ such that if
properties {\bf (S1.0)} and {\bf (S2.0)} hold for
$L_0>Q^0$, property {\bf (S1.0)} is valid  for $L_k$,
$k\geq 1$. That is, single-particle LSOs $H^{(1)}_{\Lam^{(1)}}$ satisfy
$$\begin{array}{l}
\hbox{{\bf (S1.k)} $\quad$ $\forall$ $x,y\in\Z$ and disjoint segments
 $\Lam^{(1)}_{L_k}(x)$ and $\Lam^{(1)}_{L_k}(y)$,}\\
\P\;\Big\{\forall \, E\in I:
\;{\rm{both}}\;\Lam^{(1)}_{L_k}(x )\;{\rm{and}}\;\Lam^{(1)}_{L_k}(y )
\;{\rm{are}}\;(E,m){\rm -S}\Big\} <L_k^{-2p}.
\end{array}\eqno (3.6)$$
\end{Thm}
\pmn

\myrem A detailed analysis of proofs given in \cite{vDK1} shows
that in fact, the parameters $p$ and $q$ can be chosen arbitrarily big, provided that the
amplitude
$|g|$ of the  random external potential is large enough:
$$
p \geq p(g) \tto{|g|\to\infty} + \infty,
\; q \geq q(g) \tto{|g|\to\infty} + \infty.
\eqno (3.7)
$$
\pmn

\begin{Thm}\label{vDK23} Let $I\subset\R$ be an interval of length $\leq 1$,
and fix $L_0>0$, $m>0$, $p=\vp$,
$\alpha=\valpha$ and $m>0$. Set $L_{k+1}=L_k^\alpha$, $k=0, 1, \dots$.
Suppose that for any $k=0, 1, 2, \dots$, the single-particle LSOs
$H^{(1)}_{\Lam^{(1)}}$ in {\rm{(1.7)}} obey the bound from {\bf (S1.k)}.
That is,
$$\P\;\Big\{\forall \, E\in I\;{\rm{both}}\;
\Lam^{(1)}_{L_k}(x )\;{\rm{and}}\; \Lam^{(1)}_{L_k}(y )\;
{\rm{ are}}\; (E,m){\rm -S}\Big\}\leq L_k^{-2p}, \eqno(3.8)$$
$\forall$ $x,y\in\Z$ and disjoint segments $\Lam^{(1)}_{L_k}(x)$,
$\Lam^{(1)}_{L_k}(y)$.

Then, with probability one, the spectrum of single-particle LSO $H^{(1)}$ (cf.
{\rm{(1.8)}}) in $I$ is pure point, and the EFs corresponding to EVs in
$I$ decay exponentially fast at infinity.
\end{Thm}
\pmn

The proof of Theorem \ref{vDK23} is purely deterministic and does not rely upon
probabilistic properties of the random process of potential values
$gV(x,\omega)$, $x\in\Z$. The core technical statement to be adapted
to our two-particle model is the above Theorem \ref{vDK22}. We will see, however, that
methods and results of one-particle localisation theory also play an important role in
the two-particle theory.

%

Apart from probabilistic estimates of the Green's functions in finite volumes, we will
also need the following result on the exponential decay of EFs of one-dimensional LSOs in
finite volumes. It is convenient here to introduce the definition of "tunneling".

\begin{Def}\label{defNT} Given $x\in\Z$ and an integer $L>0$,
let $\psi_j$, $j=1, \dots, 2L+1$, be the EFs of matrix
$H^{(1)}_{\Lam^{(1)}_L(x)}$, the
single-particle LSO in segment $\Lam^{(1)}_L(x)=[x-L,x+L]\cap\Z$ (cf.
{\rm{(1.7)}}). We say that
$\Lam^{(1)}_L(x)$ is $m$-non-tunneling ($m$-{\rm{NT}}, for short),
if the following inequality holds:
$$\sum_{j} \sum_{y=x\pm L}
|\psi_{j}(x)\psi_j(y)| \leq e^{-m L}.\eqno{(3.9)}$$ Otherwise, $\Lam^{(1)}_L(x)$ is
called $m$-tunneling ($m$-{\rm T}).
\end{Def}

The rest of the presentation, in Sections 3 and 4, is based on a sequence of technical
lemmas related to single- and two-particle systems.
\pmn

\begin{Lem}\label{NTplusNR} Fix $\beta =\vbeta$. Given $E\in \R$,
$x\in\Z$ and an integer $L\geq 1$, consider segment
$\Lam^{(1)}_{L}(x) = [x-L,x+L]\cap \Z$ and the single-particle
LSO $H_{\Lam^{(1)}_{L}(x)}$ in {\rm{(1.7)}}. Assume that $\Lam^{(1)}_L(x)$ is
$E$-{\rm NR} and $m$-{\rm NT} where $m\geq 2$. Then
$\Lam^{(1)}_L(x)$ is also $(E,m')${\rm -NS} where $m'$ satisfies
$$m'\geq m - L^{-(1-\beta)}.\eqno (3.10)$$
\end{Lem}
\pmn

For the proof, use the formula for the Green's functions
$G^{(1)}_{\Lam^{(1)}_L(x)}(u,y;E)$ (cf. (3.3)):
$$G^{(1)}_{\Lam^{(1)}_L(x)}(u,y;E) = \sum_{j=1}^{2L+1}
\frac{\psi_j(u){\overline\psi_j}(y)}{E_j-E},$$
where $E_j$ is the EV of the EF $\psi_j$ of
$H^{(1)}_{\Lam^{(1)}_L(x)}$.

In the one-dimensional, single-particle Anderson model,
it is well-known that the
probability of tunneling in  segment $\Lam_L^{(1)}(x)$
is exponentially small with
respect to $L$; see, e.g., \cite{GMP}, \cite{KS}). For convenience, we state here the corresponding assertion in the form used below, with a power-like bound.  In this form it has been proven in
higher dimensions, for large values of $|g|$; see \cite{vDK1}, Theorem
2.3 and Lemma 3.1. We note that in \cite{AM} a stronger bound was established,
by using the method of fractional moments of the resolvent.

\pmn

\begin{Lem}\label{ExpEF} Consider segment
$\Lam^{(1)}_L(x) = [x-L,x+L]\cap \Z$ and the LSO
$H_{\Lam^{(1)}_L(x)}$. Then for any $m>0$ there exist
constants $g_1=g_1(m)\in(0,+\infty)$ and $L_0=L_0(m)$ such that for
all $|g|\geq g_1$ and $L\geq L_0$ we have
$$\P\;\Big\{ \Lam^{(1)}_L (x)\;{\rm{is}}\;
m{\rm{-NT}}\Big\} \geq 1 - L^{-q}.
\eqno (3.11)$$
\end{Lem}
\pmn

Lemma \ref{ExpEF} plays an important
role in the proof of Lemma \ref{MainInductiveLemma}; see below. We
want to note that such strong (in fact, optimal) probabilstic
estimates, both for continuous and discrete one-dimensional random
Schr\"{o}dinger operators, go back to earlier works, viz.
\cite{GM}, \cite{GMP}, \cite{Mol}, \cite{KS}, \cite{C}.
The reader can find a detailed account of specifically one-dimensional
methods and an extensive bibliography in the monographs
\cite{CL} and \cite{PF1} (cf., in particular, Theorem VIII.3.7
and Section VIII.3 in \cite{CL}).


\section{An MSA for a two-particle system}

In this section, we propose a modification to the von Dreifus -- Klein MSA scheme so as
to adapt it to two-particle systems. The scheme allows any finite number of "singular"
areas in a given finite volume
$\Lam\subset \z2$, provided that the "disorder" is high
enough ($|g|\gg 1$). This feature is a serious advantage of the MSA scheme which makes it
flexible and applicable to the random field (1.12) generated by the potential $V(x,\om)$,
$x, x_1, x_2\in \Z^1$.

As was said before, we follow the general strategy of \cite{vDK1}, but introduce some
technical changes. It was noted that the MSA scheme includes values $p$, $q$, $\alpha$
and $\beta$ which are subject to certain restrictions. For us, it is convenient to set,
throughout Sections 4 and 5:
$$p=\vp,\;\;q=\vq,\;\;\alpha =\valpha,\;\text{ and }\;\beta =\vbeta,
\eqno (4.1) $$
similarly to Section 3. However, to make the presentation consistent with that in
\cite{vDK1}, we continue referring to parameters $p$,
$q$, $\alpha$ and $\beta$ in our constructions below. The main
components of the MSA scheme are an increasing sequence of positive integer lengths
$L_0,L_1, L_2,
\dots$ and a decreasing sequence of positive masses
$m_0, m_1, m_2,\dots$. In Sections 4 and 5 these sequences are assumed
to be as follows:

(i) for $k\geq 1$:
$$L_k = \text{ the smallest integer
$\geq L_{k-1}^\alpha$}\eqno (4.2)$$
and
$$
m_k = m_0 \, \prod_{j=1}^k \left( 1 - 8 L_0^{-j/2} \right),\eqno (4.3)$$

(ii) $L_0$ is positive integer and $m_0$ is positive such that
$$ m_0>2,\;\;
L_0\geq 256 \text{ and } \, e^{-m_0L_0} \leq e^{-L_0^\beta}.\eqno (4.4)$$

Observe that, owing to the bound $L_0 \geq 256$, the infinite product\\
$\diy{\prod\limits_{j\geq 1}}(1-8L_0^{-j/2}) \geq 1/2$. Thus,
$$m_\infty:=\liminf_{k\to\infty} m_k \geq m_0/2. \eqno (4.5)$$
In addition, we will have to assume that $L_0$ is large enough; such restrictions will
apeear in various lemmas below. Ultimately, the lower bound on
$L_0$ will depend on a particular choice of $m_0$.

In this respect, it should be noted that the choice of $m_0$ and $L_0$ dictates the
choice of the value of $|g|$ in Theorem 1.1. More precisely, if $|g|$ large enough
(roughly, $\ln\;|g|\simeq O(m_0L_0)$), then the (modified) MSA scheme will guarantee the
exponential decay of the EFs of the two-particle LSO $H$ from (1.2) with mass $\geq
m_\infty$.
\pmn

As in \cite{vDK1}, we define the notions of resonant and singular sub-squares.

\begin{Def} Given $E\in\R$,
a sub-square $\Lam_L(\ux )$ of size $L$ with center at
$\ux $ is called $E$-resonant
($E${\rm -R}, for short) if the spectrum
$\sigma(H^{(2)}_{\Lam_L(\ux )})$
of $H^{(2)}_{\Lam_L(\ux )}$, the two-particle LSO in $\Lam_L(\ux )$, satisfies
$$
{\rm dist}\left[E, \sigma\big(H^{(2)}_{\Lam_L(\ux )}\big)\right]
< e^{-L^\beta}.\eqno (4.6)$$
Otherwise, $\Lam_L(\ux )$ is called $E$-non-resonant ($E${\rm{-NR}}).
\end{Def}

\begin{Def} Given $E\in\R$ and $m>0$,
a sub-square $\Lam_L(\ux )$ is called $(E,m)$-non-singular
($(E,m)${\rm -NS}, for short), if
$$
\max_{\uu\in \pt \Lam_L(\ux )}\left|\,G^{(2)}_{\Lam_L(\ux )}
(\ux ,\uu ; E; \om)\right| \leq e^{-mL}.\eqno (4.7)$$
Otherwise it is called $(E,m)$-singular ($(E,m)$-S). Here,
$G^{(2)}_{\Lam_L(\ux )}(\uy ,\uu ; E)$, $\uy,\uu\in \Lam_L(\ux)$,
stands for the Green's function of $H^{(2)}_{\Lam_L(\ux)}$:
$$
G^{(2)}_{\Lam_L(\ux )}(\uy ,\uu ;E) = \left\langle \left( H^{(2)}_{\Lam_L(\ux )}- E
\right)^{-1} \delta_\uy, \delta_\uu
\right\rangle, \;\;\uy,\uu\in \Lam_L(\ux).
\eqno (4.8)
$$
\end{Def}

Similar definitions hold for $H^{\rm{n-i}}_{\Lam_L(x_1)\times
\Lam_L(x_2)}$, the LSO of a two-particle system with
no interaction, in square $\Lam_L(x_1)\times\Lam_L(x_2)$; cf.
(1.9).

Recall, in this paper, the interaction potential $U$ has finite range $d$ (cf. (1.4)).
So, there are two kinds of sub-squares: those which are disjoint from the diagonal strip
$$
\cD=\myset{\uz=(z_1,z_2)\in\z2_{\geq}:\,z_1-z_2\leq d },
$$
and those having common points with $\cD$. The former are called off-diagional
sub-squares (actually, squares), and the latter diagonal sub-squares. On an off-diagonal
square $\Lam_{l}(\uy )$, the interaction potential is identically zero, and so the
two-particle LSO
$H^{(2)}_{\Lam_{l}(\uy )}$ coincides with
$H^{\rm{n-i}}_{\Lam_{l}(\uy )}$
(cf. (1.9)) and is written as the sum (1.11) involving single-particle LSOs
$H^{(1)}_{\Lam^{(1)}_l(y_1)}$ and $H^{(1)}_{\Lam^{(1)}_l(y_2)}$. The
distinction between off-diagonal and diagonal sub-squares requires different techniques.

Our version of the two-particle MSA scheme can be summarised in a form similar to that in
Section 3. More precisely, the following assertions hold, whose structure is similar to
Theorems \ref{vDK22} and \ref{vDK23}:
\pmn

\begin{Thm}\label{Thrm4.1} Let $I\subset\R$ be an interval of length $\leq 1$.
Given $L_0>0$, $m_0>0$, consider the following properties {\bf (T1.0)} and
{\bf (T2.0)} of two-particle LSOs
$H_{\Lam_L}$ from {\rm{(1.5)}}:
\psn
$$\begin{array}{l}\hbox{{\bf (T1.0)} $\quad$
$\forall$ $\ux,\uy\in\z2_{\geq}$ and $L_0$-{\rm D}
sub-squares $\Lam_{L_0}(\ux)$ and $\Lam_{L_0}(\uy)$,}\\
\P\;\Big\{\forall \, E\in I:\;{\rm{both}}\;\Lam_{L_0}(\ux)
\;{\rm{and}}\;\Lam_{L_0}(\uy)
\;{\rm{are}}\;(E,m_0){\rm -S}\Big\}<L_0^{-2p}.
\end{array}\eqno (4.11)$$
\psn
$$\begin{array}{l}\hbox{{\bf (T2.0)} $\quad$ $\forall$ $L\geq L_0$
and $\forall$ $E$ with
${\rm dist}\left[E,I\right]\leq\half e^{-L^\beta}$,}\\
\P\;\Big\{
 \; \Lam_L(\ux) \text{ is } E\text{\rm-R}
\Big\}< L^{-q}.\end{array}\qquad\quad\;\eqno (4.12)$$
\psn

Next, define values $L_k$ and $m_k$, $k\geq 1$, as in {\rm{(4.2)}} and
{\rm{(4.3)}}. There exists $Q_0=Q_0(m_0)\in (0,\infty)$ such that if
properties {\bf (T1.0)} and {\bf (T2.0)} are valid for $L_0>Q_0$ and
$m_0$, then property {\bf (T1.0)} holds for $L_k$ and $m_k$, $k\geq 1$.
That is, the two-particle LSOs $H_{\Lam_{L_k}}$, $k=1,2,\ldots$, obey
$$\begin{array}{l}
\hbox{{\bf (T1.k)} $\quad$ $\forall$ $\ux,\uy\in\Z_{\geq}$ and $L_k$-{\rm D}
sub-squares $\Lam_{L_k}(\ux)$ and $\Lam_{L_k}(\uy)$,}\\
\P\;\Big\{\forall \, E\in I:
\;{\rm{both}}\;\Lam_{L_k}(\ux)\;{\rm{and}}\;\Lam_{L_k}(\uy)
\;{\rm{are}}\;(E,m_k){\rm -S}\Big\} <L_k^{-2p}.
\end{array}\eqno (4.13)$$
\end{Thm}
\pmn

\begin{Thm}\label{Thrm4.2} Let
$I\subset\R$ be an interval of length $\leq 1$, and fix
$L_0>0$, $m_0>0$. Define values $L_k$, $k\geq 1$, as in {\rm{(4.2)}},
and suppose that for some finite constant $C$, for any
$k=0, 1, 2, \dots$, two-particle LSOs $H_{\Lam_L}$
from {\rm{(1.5)}} satisfy the bound
$$
\P\;\Big\{\forall\;E\in I:\;{\rm{both}}\;
\Lam_{L_k}(\ux )\;{\rm{and}}\;\Lam_{L_k}(\uy )
\;{\rm{are}}\;(E,m_\infty){\rm -S}\Big\}\leq C\,L_k^{-2p}\eqno(4.14)
$$
whenever  $\ux,\uy\in\Z^2_\geq$ and $\Lam_{L_k}(\uy )$,
$\Lam_{L_k}(\uy )$ are $L_k$-distant  sub-squares. Here,
$m_\infty$ is defined in {\rm{(4.3)}} and {\rm{(4.5)}}. Then, with
probability one, the spectrum of the two-particle LSO $H$ (cf. {\rm{(1.2)}}) in $I$ is
pure point, and the EFs corresponding to the EVs in the interval
$I$ decay exponentially fast at infinity, with mass $\geq m_\infty$.
\end{Thm}
\pmn

As in Section 3, the assumption that $I$ has length $\leq 1$ is introduced for technical
convenience and does not restrict generality.

However, the reader should note a difference between Theorem \ref{vDK22} and Theorem
\ref{Thrm4.1}. Namely, in Equations (4.11) and (4.13), sub-squares
$\Lam_{L_0}(\ux)$, $\Lam_{L_0}(\uy)$ and $\Lam_{L_k}(\ux)$,
$\Lam_{L_k}(\uy)$, are assumed to be not simply disjoint but $L_0$-D
and $L_k$-D, respectively. In other words, in the two-particle MSA inductive scheme from
this paper it is required less and assumed less compared with the single-particle one
from [9]. Formally, the original argument developed in [9] estimates, at each inductive
step, the probability that any disjoint pair of volumes (cubes) is simultaneously
singular or simultaneously resonant, is sufficiently small. However, a careful analysis
shows that it suffices to consider pair of volumes satisfying a stronger requirement:
$$d_\infty ( \Lam_{L_k}(\ux), \Lam_{L_k}(\uy)) \geq C L_k,$$
for any given positive constant $C$.

The initial step in the inductive scheme described in Theorems
\ref{Thrm4.1} and \ref{Thrm4.2} is provided by Theorem \ref{Thrm4.3}
below.
\pmn

\begin{Thm}\label{Thrm4.3} $\forall$ $L_0\geq 256$ $\exists$
$g_0\in (0,\infty)$ such that, for $g$ with $|g|\geq g_0$,
$\forall$ interval $I$ of length $\leq 1$ assumptions
{\bf (T1.0)} and {\bf (T2.0)} hold true.
\end{Thm}
\pmn

Theorem 1.1 follows directly from Theorems \ref{Thrm4.1} --
\ref{Thrm4.3}. In turn, Theorem \ref{Thrm4.3} follows from
Lemmas \ref{BasicLemma1} and \ref{BasicLemma2} below.
\pmn

\begin{Lem}\label{BasicLemma1} Given $m,L>0$ and $E\in\R$,
assume that sub-square $\Lam_{L}(\ux )$ is $(E,m)${\rm{-S}}. Then $\Lam_{L}(\ux )$
contains at least one site
$\uu =(u_1,u_2)$, with
$$
|U(\uu )+gV(u_1,\om)+gV(u_2,\om)- E | < e^{mL} + \|H^0\|
=  e^{mL} + 4.\eqno (4.15)$$
Therefore, $\forall\, \tilde p >0$
$$
\begin{array}{l}
L^{\tilde p} \cdot  \P\Big\{\Lam_L(\ux)\;{\rm{is}} \;(E,m)-{\rm S}\Big\}
 \tto{|g|\to \infty} 0.
\end{array}$$

Here $\|H^0\|$ stands for the operator norm of $H^0$.
\end{Lem}
\pmn

Therefore, for any $\tilde p>0$ and all sufficiently large $|g|$,
$$
\P\Big\{\Lam_L(\ux)\;{\rm{is}} \;(E,m)-{\rm S}\Big\}
\leq L^{-\tilde p}.
$$

\pmn

\begin{Lem}\label{BasicLemma2} Given $E\in\R$,
$\forall$  $\ux\in\Z^2$ and $L\geq 1$,
$$\P\;\Big\{\Lam_{L}(\ux)\;{\rm{is}}\;E\text{{\rm -R}}\Big\}
\leq |\Lam_{L}(\ux)|^2 \,\|f\|_\infty \, e^{-L^{\beta}}.
\eqno (4.16)$$
Here and below, $|\Lam_L(\ux )|$ stands for the number of points in sub-square
$\Lam_L(\ux )$ (which is $\leq (2L+1)^2$), and $\|f\|_\infty$,
as before, is the sup-norm of PDF $f$.

Therefore, for $L>0$ large enough,
$$
\P\;\Big\{ \Lam_{L}(\ux)\;{\rm{is}}\;E\text{{\rm -R}}\Big\} \leq L^{-q}.
\eqno (4.17)$$
\end{Lem}
\pmn

Lemmas \ref{BasicLemma1} and \ref{BasicLemma2} follow directly from our Wegner-type
estimate in Theorem 2.1 (cf. Theorem A.1.3(i) in \cite{vDK1}). The meaning of this lemma
is that if a finite (and fixed) size
sub-square $\Lam_L(\ux )$ is singular and the coupling constant $g$ is large enough, then
$\Lam_L(\ux )$ contains necessarily resonant points. The importance of such a relation
between resonant and singular domains is explained by the fact that the probability of
being resonant is much simpler to estimate than that of being singular.

Both Lemma \ref{BasicLemma2} and Lemma \ref{BasicLemma2}  do not use the recursive
scheme from (4.2) and (4.3).
\pmn

The estimates provided by Lemma \ref{BasicLemma1} and Lemma \ref{BasicLemma2} will also
be used in the proof of Lemma \ref{MainInductiveLemma}, in the same way as a similar
estimate was used in \cite{vDK1}.

The statement of Theorem \ref{Thrm4.2} is similar to the assertion of Lemma 3.1 from
\cite{vDK1}. We want to note that Lemma 3.1 in
\cite{vDK1} is a general statement based only on probabilistic
estimates provided by Lemma 3.2 in \cite{vDK1}, so that the Borel-Cantelli lemma (which
is the key ingredient of the proof of Lemma 3.1 in \cite{vDK1}), applies. In our
situation, the proof of Theorem \ref{Thrm4.2} goes along the same line and is based on
probabilistic estimates from Theorem \ref{Thrm4.1}.

Therefore, to prove Theorem 1.1, it suffices to establish Theorem
\ref{Thrm4.1}. This is the subject of the rest of the paper. The
specification (4.1) will help with producing fairly explicit bounds. In essence, Theorem
\ref{Thrm4.1} constitutes an inductive assertion, in the value of $k$, guaranteeing a
reproduction of property
{\bf (T1.$k$)} from properties {\bf (T1.$k-1$)} and {\bf (T2.$0$)}.
More precisely, in our approach to inequality (4.13), we estimate (by using different
methods), the probability in the LHS of (4.13) for pairs of $L_k$-D sub-squares
$\Lam_{L_k}(\ux)$ and $\Lam_{L_k}(\uy)$ of three types:
\psn

(I) Both $\Lam_{L_k}(\ux)$
and $\Lam_{L_k}(\uy)$ are off-diagonal; see Lemma \ref{TwoOffDiagSing}.
\psn

(II) Both $\Lam_{L_k}(\ux)$ and $\Lam_{L_k}(\uy)$ are diagonal;
 see Lemmas \ref{ProbDiagonalSing}
and \ref{MainInductiveLemma}.
\psn

(III) One of sub-squares $\Lam_{L_k}(\ux)$ and $\Lam_{L_k}(\uy)$ is
off-diagonal and one diagonal; see Lemma \ref{DiagAndOffdiag}.
\pmn

As was said earlier, the aim is to show that the probability figuring in {\bf (T1.$k$)},
that $\forall\;E\in I$, both $
\Lam_{L_k}(\ux )$ and $\Lam_{L_k}(\uy )$
are $(E,m_k)$-S, in each of cases (I) -- (III) is bounded from above by $L_k^{-2p}$. Here
and below,
$L_k$ and $m_k$ are assumed to be as in (4.2) and (4.3).
\pmn

The rest of Section 4 contains various bounds on probabilities related to LSO
$H_{\Lam_L(\ux )}$. The first such bound is
\pmn

\begin{Lem}\label{DirectSum} Fix $m>2$. There exists
$Q_1=Q_1(m)$ such that for $L\geq Q_1$ the following property holds.
Let $\ux\in\z2_{\geq}$ and assume that $\Lam_L(\ux
)=\Lam^{(1)}_L(x_1)\times\Lam^{(1)}_L(x_2)$ is an off-diagonal square. Assume that
segments
$\Lam^{(1)}_L(x_1)$ and $\Lam^{(1)}_L(x_2)$
are $(E,m)${\rm -NT}, in the sense of Definition {\rm \ref{defNT}}. Consider the LSO
$H_{\Lam_L(\ux )}=H^{\rm{n-i}}_{\Lam_L(\ux )}$. Then there exists positive $m'$
satisfying
$$m'\geq m - 3 L^{-(1-\beta)}   \eqno (4.18)$$
such that if  square $\Lam_L(\ux )$ is $E${\rm -NR}, then it is
$(E,m')${\rm -NS}.
\end{Lem}
\pmn

For the proof of Lemma \ref{DirectSum}, see Section 5.
\pmn

The next assertion, Lemma \ref{TwoResSubSquares}, helps to understand several parts of
the two-particle MSA scheme. Consider standard coordinate projections $\Pi_j: \z2\to\Z$,
$j=1,2$,
so that, for a given subset of the lattice $\Lam\subset \z2$, its coordinate projections
are given by
$$\Pi_1(\Lam ) = \myset{ u_1\in\Z: \;(u_1,u_2)\in\Lam\;\hbox{ for
some $u_2\in\Z$} },$$ and
$$\Pi_2(\Lam )=\myset{u_2\in\Z: \;(u_1,u_2)\in A\;\hbox{ for
some $u_1\in\Z$}}.$$
\pmn

\begin{Lem}\label{TwoResSubSquares} Fix an interval $I\subset\R$ of length
$\leq 1$. Suppose that property {\bf{(T2.0)}} holds, that is
$$
\P\;\Big\{
 \; \Lam_L(\ux) \text{ is } E\text{\rm-R}
\Big\}< L^{-q},\;
\hbox{$\forall$ $L\geq L_0$ and $E$ with
${\rm dist}\left[E,I\right]\leq\half e^{-L^\beta}$.}
$$
Next, let $\Lam'=\Lam_{L'}(\uu')$ and
$\Lam''=\Lam_{L''}(\uu'')$ be two sub-squares such that
both their horizontal projections are disjoint and their vertical projections are
disjoint:
$$\Pi_1(\Lam')\cap\Pi_1(\Lam'')=\emptyset,
\;\;
\Pi_2(\Lam') \cap \Pi_2(\Lam'')=\emptyset.
$$
Set $L = \min\{L',L''\}$. If $L\geq L_0$, then
$$\P\;\Big\{\exists E\in I:\;{\rm{both}}\;\Lam'\;{\rm{and}}\;
\Lam''\;{\rm{are}}\;E{\rm -R}\Big\}\leq L^{-q}.\eqno (4.19)$$
\end{Lem}
\pmn

The proof of Lemma \ref{TwoResSubSquares} is straightforward; see Section 5. Observe that
Lemmas \ref{BasicLemma1} -- \ref{TwoResSubSquares} are "non-recursive" statements: they
do not refer the recursive scheme introduced in (4.2) and (4.3).

We now pass to Lemma \ref{TwoOffDiagSing} which covers the probability in the LHS of
(4.13) for two off-diagonal squares $\Lam_{L_k}(\ux )$ and $\Lam_{L_k}(\uy )$. The
estimate provided in this lemma is similar to that in Lemma 4.1 from \cite{vDK1}.
However, the difference is that in  Lemma \ref{TwoOffDiagSing} the assumption is made for
all pairs of disjoint sub-squares and reproduced for pairs of disjoint off-diagonal
squares.
\pmn

\begin{Lem}\label{TwoOffDiagSing} Let $I\subset\R$ be an
interval of length $\leq 1$. Suppose that, $\forall\,\, L\geq L_0$, property
(3.12) is fulfilled. That is:
$$
\P\;\Big\{ \Lam^{(1)}_L (x)\;{\rm{is}}\; m{\rm{-NT}}\Big\} \geq 1 - L^{-q}.
$$
Then, $\forall$ pair of $L_k$-{\rm D}, off-diagonal squares $\Lam_{L_{k+1}}(\ux)$ and
$\Lam_{L_{k+1}}(\uy)$,
$$\pr{\exists\;E\in I:\,\text{ both }\Lam_{L_{k+1}}(\ux)
\text{ and } \Lam_{L_{k+1}}(\ux)
\text{ are } (E,m_{k+1})-{\rm S} } \leq L_{k+1}^{-2 p}.\eqno (4.20)
$$
\end{Lem}
\pmn

For the proof of Lemma \ref{TwoOffDiagSing}, see Section 5.
\pmn

The assertion of Lemma \ref{AnalyticResolvent} below is close to Lemma 4.2 in
\cite{vDK1} and can be proved in essentially the same way, for
it only relies upon singularity/non-singularity properties of the sub-squares residing in
a larger sub-square.
\pmn

\begin{Lem}\label{AnalyticResolvent}
Fix $E\in\R$ and an integer $K>0$. There exists a constant
$Q_2 = Q_2(K)\in(0,+\infty)$ with the following property. Assume that
$L_0\geq Q_2$. Next, given $k\geq 0$, assume that a sub-square
$\Lam_{L_{k+1}}(\ux)$ is $E${\rm-NR} and does not contain more than
$K$ disjoint sub-squares $\Lam_{L_k}(\uu_i)\subset \Lam_{L_{k+1}}(\ux)$
that are $(E,m_k)$-{\rm S}. Then  sub-square $\Lam_{L_{k+1}}(\ux)$ is
$(E,m_{k+1})$-{\rm{NS}}.
\end{Lem}
\pmn

Next, consider an assertion
\pmn\DSk: {\sl $\forall$ pair of  diagonal $L_k${\rm{-D}} sub-squares
$\Lam_{L_k}(\uu )$ and $\Lam_{L_k}(\uv )$}
$$
\P\;\Big\{ \exists \, E\in I:\,{\rm{both}}\;
\Lam_{L_k}(\uu )\;{\rm{and}}\;\Lam_{L_k}(\uv )\;{\rm{are}}
\;(E,m_k)\text{-S}\Big\}\leq L_k^{-2p}.
\eqno (4.21)$$
The following lemma is used in the proof of Lemma
\ref{MainInductiveLemma}.
\pmn

\begin{Lem}\label{ProbDiagonalSing}
Given $k\geq 0$, assume that property \DSk in
{\rm{(4.21)}} holds true. Consider a \sq $\Lam:=\Lam_{L_{k+1}}(\ux )$
and let $N(\Lam;E)$ be the maximal number of $(E,m_k)${\rm -S}, pair-wise $L_k${\rm{-D}}
diagonal sub-squares
$\Lam_{L_k}(\uu^{(j)})\subset \Lam$. Then $\forall$ $n\geq 1$,
$$
\P\;\Big\{\exists E\in I:\;\; N(\Lam;E) \geq 2n\Big\}\leq
L_k^{n(1+\alpha)} \cdot  L_k^{-np/2}.\eqno (4.22)$$
\end{Lem}
\pmn

Now comes a statement which extends Lemma 4.1 from \cite{vDK1} to pairs of diagonal
sub-squares.
\pmn
\begin{Lem}\label{MainInductiveLemma}
There exists a constant $Q_3\in(0,+\infty)$ such that if $L_0\geq Q_3$, then, $\forall\,
k\geq 0$, the property \DSk $\,$ in {\rm{(4.21)}} implies \DSkone .
\end{Lem}
\pmn

Finally, the case of a pair with one diagonal and one off-diagonal sub-square is covered
by
\pmn
\begin{Lem}\label{DiagAndOffdiag} There exists a constant
$Q_4\in(0,+\infty)$ with the following property. Assume that
$L_0\geq Q_4$ and that, given $k\geq 0$, property \DSk in {\rm{(4.21)}}
holds. Let
$\Lam'=\Lam_{L_{k+1}}(\ux')$ be a diagonal sub-square and
$\Lam'' = \Lam_{L_{k+1}}(\ux'')$ an off-diagonal square, and let
$\Lam'$ and $\Lam''$ be $L_{k+1}$-{\rm D}. Then
$$
\P\;\Big\{\exists \, E\in I:\,  \text{\rm both } \Lam'\;{\rm{and}}\;
\Lam''\;{\rm{are}}\;(E,m_{k+1}){\rm -S}\Big\}\leq L_{k+1}^{-2p}.
\eqno (4.23)$$
\end{Lem}
\pmn

From Lemmas \ref{TwoOffDiagSing}, \ref{MainInductiveLemma}, and
\ref{DiagAndOffdiag}, Theorem \ref{Thrm4.1} is deduced by following
the remaining parts of the MSA scheme \cite{vDK1}.
\pmn

\section{Proof of MSA Lemmas}
\pmn
\myproof{Lemma \ref{DirectSum}}
Let $\{\psi'_j\}$ be normalised EFs of single-particle LSO
$H^{(1)}_{\Lam^{(1)}_L(x_1)}$
with EVs $E'_j$ and $\{\psi''_k\}$ be normalised eigen-functions of
$H^{(1)}_{\Lam^{(1)}_L(x_2)}$ with EVs $E''_k$.  As $\norm{\psi'_j}_2 = \norm{\psi''_k}_2
= 1$, we have that
$$\max_{u\in\Lam^{(1)}_L(x_1)}\abs{ \psi'_j(u) } \leq 1,
\;\max_{v\in\Lam^{(1)}_L(x_2)} \abs{ \psi''_k(v) } \leq 1.$$

Next, for $\uu=(u,u'),\uv=(v,v')\in\Lam_L(\ux )$, the two-particle Green's functions have
the form
$$\begin{array}{cl}
G_{\Lam_{L}(\ux)}(\uu,\uv ;E)&=
\left\langle \left(
H^{({\rm{n-i}})}_{\Lam_L(\ux )}- E \right)^{-1} \delta_\uu, \delta_\uv
\right\rangle\\
\;&=\sum\limits_{j,k}\;
\diy{\frac{\psi'_j(u'){\ol\psi'_j(v')}
\psi''_k(u''){\ol\psi''_k(v'')}}
{E- (E'_j + E''_k)}}.\end{array}
\eqno (5.1)$$

Further, assuming that $\Lam_{L}(\ux )$ is $E${\rm -NR}, we get
$$\abs{E- (E'_j + E''_k)}^{-1} \leq e^{L^\beta}.$$
Finally, for $\uy\in\pt\Lam_{L,L'}(\ux )$,
$$
\begin{array}{cl}\abs{ G_{\Lam_{L,L'}(\ux )}(\ux, \uy; E) }
&\leq (2L+1)^2 \;\frac{\displaystyle{e^{-mL}}}
{ \displaystyle{\min_{j,k}  \abs{ E - (E^1_j+E^2_k}}}\\
\;&\leq (2L+1)^2 e^{-mL} e^{L^\beta} \leq e^{-m' \, L},
\end{array}\eqno (5.2)$$
with
$$m' = m - L^{-1} \left(2\,\ln\,(2L+1) + L^\beta \right)
\geq m - C' L^{-(1-\beta)}.\qed\eqno (5.3)$$
\pmn

\myproof{Lemma \ref{TwoResSubSquares}}\;Since
$\Lam'$ and $\Lam''$ have both coordinate projections disjoint, the
respective samples of potential in these two sub-squares are independent, as in the
single-particle theory with IID potential. So, we can use exactly the same argument
(conditioning on the potential in $\Lam'$, combined with the Wegner-type estimate for a
single-particle model) as in the proof of Lemma 4.1 in \cite{vDK1}. $\qquad\qed$
\pmn

\myproof{Lemma \ref{TwoOffDiagSing} }
If $\Lam'$ and $\Lam''$ are off-diagonal and $L_k$-D, then, by virtue of Lemma 2.4,  at
least one of their horizontal and vertical projections among
$$
I_1=\Pi_1\Lam',\;J_1=\Pi_2\Lam',\;I_2=\Pi_1\Lam'',\;J_2 =\Pi_2\Lam''
$$
is disjoint with the three others. Without loss of generality, suppose that
$$
I_1\cap (J_1\cup I_2\cup J_2) = \emptyset;
$$
three other possible cases are similar.

Consider the following events:
$$\begin{array}{l}
B=\myset{ \text{ both }\Lam'\text{ and  }\Lam''\text{ are }
(E,m_{k+1})\text{-S }},\\
C=\myset{ \text{ both }\Lam'\text{ and  }\Lam''\text{ are } E\text{-R }}.\end{array}\eqno
(5.4)$$ Then we can write
$$\pr{C}=\E\Big[{\pr{C \,|\,{\mathfrak V}(
\Pi_1\Lam'\cup\Pi_2 \Lam'')}}\Big].$$
where the sigma-algebra ${\mathfrak V}(
\Pi_1\Lam'\cup\Pi_2\Lam'')$ is generated by potential values $\,$
$\myset{V(x,\cdot),\,x\in\Pi_1\Lam'\cup\Pi_2 \Lam''}$.

By Theorem 2.3, the conditional probability $\pr{C\,|\,
{\mathfrak V}(\Pi_1\Lam'\cup\Pi_2\Lam'')}$
is a.s. bounded by $L_{k}^{-q}$, and so is its expectation.

Now let
$$\tilde J=\Pi_1\Lam'\cup\Pi_1\Lam''\cup\Pi_2\Lam'\cup\Pi_2\Lam''
$$
and consider the event
$$T=\myset{\exists\,\text{ an }(E,m_0+1)\text{-T interval }J_{L_{k+1}}
\subset \tilde J }.\eqno (5.5)$$
By Lemma 3.2 and Corollary 3.1, $\pr{T} \leq L_{k+1}^{-q}$. On the other hand, if the
potential sample belongs to $\bar T$ and both $\Lam'$ and $\Lam''$ are $(E,m_{k+1})$-S,
then both sub-squares must be $E$-R.

Now we can write that
$$\begin{array}{cl}
\pr{B}&\leq\pr{B\cap\bar C\cap\bar T)}+\pr{B\cap C}+\pr{B \cap T}\\
\;&\leq \pr{B\cap\bar C\cap\bar T} + \pr{C} + \pr{T}.\end{array}
$$
By Lemma 4.3, if $\Lam'$ (resp., $\Lam''$) is both $(E,m_{k+1})$-NT
(property $\bar T$) and $E$-NR (property $\bar C$), then it cannot
be $E$-R, so that $B\cap\bar C\cap\bar T =\emptyset$. Finally,
$$\pr{B}\leq\pr{B\cap C}+\pr{B\cap T}<L_{k+1}^{-q}+L_{k+1}^{-q}
= 2L_{k+1}^{-q}< L_{k+1}^{-2p}.\qed\eqno (5.6)$$
\noindent
\pmn

\myproof{Lemma \ref{ProbDiagonalSing}}\;Suppose
we have diagonal sub-squares $\Lam_{L_k}(\uu^{(1)})$, $\ldots$,
$\Lam_{L_k}(\uu^{(2n)})$, such that
\psn
a) any two of them are $L_k$-D, i.e., are at the distance
$\geq 6L_k+2d$,
\psn
b) all the sub-squares $\Lam_{L_k}(\uu^{(1)})$, $\ldots$,
$\Lam_{L_k}(\uu^{(2n)})$ lie in
$\Lam=\Lam_{L_{k+1}(\ux)}$.

\noindent Without loss of generality, one
can assume that points $\uu^{(i)}=\left(u_1^{(i)},u_2^{(i)}\right)$ have
$$u^{(1)}_1 < u^{(2)}_1 < \dots < u^{(2n)}_1.$$
Indeed, one can always sort entries $u^{(i)}_1$ in the non-decreasing order, and if two
of them coincide, say
$u^{(j)}_1 = u^{(j+1)}_1$, then sub-squares $\Lam_{L_k}(\uu^{(j)})$
and $\Lam_{L_k}(\uu^{(j+1)})$ cannot be disjoint, which is impossible by our hypothesis.
Then it is readily seen that:

(i) By virtue of Lemma 2.3, $\forall$ pair $\Lam_{L_k}(u^{(j)})$, $\Lam_{L_k}(u^{(j+1)})$,
the respective (random) LSOs
$H_{\Lam_{L_k}(\uu^{(j)})}$ and $H_{\Lam_{L_k}(\uu^{(j+1)})}$1
are independent, and so are their spectra and Green's functions.

(ii) Moreover, the pairs of LSOs,
$$\left(H_{\Lam_{L_k}(\uu^{(2j+1)})},H_{\Lam_{L_k}(\uu^{(2j+2)})}
\right),\;\;j=0, \dots, n-1,$$
form an independent family. Thus, any collection of events $\cA_0$,
$\ldots$, $\cA_{n-1}$ related to the corresponding pairs
$\left(H_{\Lam_{L_k}(\uu^{(2j+1)})},H_{\Lam_{L_k}(\uu^{(2j+2)})}
\right)$, $j=0, \ldots,$ $n-1$, also form an independent family.

Indeed, LSO $H_{\Lam_{L_k}(\uu^{(j)})}$ is measurable with respect to the sigma-algebra
${\mathfrak V}(I_j\cup J_j)$ generated by random variables
$V(u,\cdot)$, $u\in \,I_j \bigcup J_j$, where
$$I_j=\Lam^{(1)}_{L_k}(u_1^{(j)})=\Pi_1(\Lam_{L_k}(\uu^{(j)})),\,
J_j=\Lam^{(1)}_{L_k}(u_2^{(j)})=\Pi_2(\Lam_{L_k}(\uu^{(j)})).$$ By virtue of Lemma 2.3,
sigma-algebras ${\mathfrak V}(I_j\cup J_j)$, $j=1,\dots, 2n$, are independent. Then the
sigma-algebras $${\mathfrak V}(I_{2j+1}
\cup J_{2j+1})\vee {\mathfrak V}(I_{2j+2}\cup J_{2j+2}),\;\;j=0, \dots, n-1,$$
generated by subsequent pairs are also independent,

Now, for $j=0,\ldots, n-1$,  set
$$\cA_j = \myset{ \exists \, E\in I:\,\Lam_{L_k}(\uu^{(2j+1)})
\;{\rm{and}}\;\Lam_{L_k}(\uu^{(2j+2)})\;{\rm{are}}\;
(E,m_k)\text{-S} }.\eqno (5.7)$$
\psn

Then, by the hypothesis \DSk,
$$\P\;\Big\{ \cA_j\Big\}\leq L_k^{-p},\;\;0\leq j\leq n-1,
\eqno (5.8)$$
and by virtue of independence of events $\cA_0$,
$\ldots$, $\cA_{n-1}$, we obtain
$$\P\;\Big\{ \bigcap_{j=0}^{n-1} \cA_j\Big\} =
\prod_{j=0}^{n-1}\P\;\Big\{\cA_j\Big\}\leq
\left(L_k^{-q}\right)^{n}.\eqno (5.9)$$
To complete the proof, it suffices to notice that the total number of different families
of $2n$ sub-squares with required properties is bounded by
$2 \cdot L_k\cdot L_{k+1}$, since their centres must belong
to a strip $\myset{(x_1,x_2)\in\Z^2:\,x_1\leq x_2+2L_k} \cap \Lam_{L_{k+1}}$ of width
$2L_k$ adjoint to the diagonal $\partial\z2_{\geq}$.
$\qquad\qed$
\pmn
\myproof{Lemma \ref{MainInductiveLemma}}
The strategy of this proof is very close in spirit to that of Lemma 4.1 in \cite{vDK1}.
This similarity is due to a simple geometrical fact: samples of potential corresponding
to two
$L_{k+1}$-D diagonal sub-squares $\Lam'$, $\Lam''$ of size $L_{k+1}$
are independent, for both their horizontal projections are disjoint and their vertical
projections are disjoint. This makes the situation quite similar to that of lemma 4.1 in
\cite{vDK1}. The difference, though, is that inside each of
the sub-squares, smaller scale sub-squares $\Lam_{L_k}(\uu)$ are not pair-wise
independent, so we need to use a more involved proof based on our conditional Wegner-type
estimates.

Let $k\in \N$ and $\Lam_{L_k}(\uu)$, $\Lam_{L_k}(\uv)$ be two diagonal $L_k${\rm{-D}}
sub-squares. Consider the following event:
$$B_k(\uu, \uv)
= \Big\{ \exists \, E\in I:\,{\rm{both}}\;
\Lam_{L_k}(\uu )\;{\rm{and}}\;\Lam_{L_k}(\uv )\;{\rm{are}}
\;(E,m_k)\text{-S}\Big\}\,.\eqno (5.10)$$
Assuming that the estimate
$$\pr{ B_k(\uu,\uv)} \leq L_k^{-2p}\eqno (5.11)$$
holds for all diagonal $L_k${\rm{-D}} sub-squares $\Lam_{L_k}(\uu)$,
$\Lam_{L_k}(\uv)$, we have to obtain the similar estimate at
scale $L_{k+1}$. Namely, fix two diagonal and
$L_{k+1}${\rm{-D}} sub-squares, $\Lam_{L_{k+1}}(\uu)$
and $\Lam_{L_{k+1}}(\uv)$. Then we have to prove that
$$\pr{ B_k(\uu,\uv)} \leq L_{k+1}^{-2p}.\eqno (5.12)$$

We will do so by covering the event $B_k(\uu,\uv)$ by a union of several events the
probability of which will be estimated separately. We shall use shortened notations
$B_k=B_k(\uu,\uv)$, $\Lam'= \Lam_{k+1}(\uu)$,
$\Lam''= \Lam_{k+1}(\uv)$.

It is convenient to introduce three events:
\pmn
$$\begin{array}{cl}
C &=\{\text{ either }\Lam'\text{ or }\Lam''\text{ contains at least }2n,\,\,
(E,m_k)\text{-S, }\\
\;&\quad\;\;\text{ pair-wise } L_k\text{-D }\text{ sub-squares }
\Lam_k(\uu_i),\;i=1,\ldots,2n\},\\
D&=\{\text{ either }\Lam'\text{ or }\Lam''\text{ contains at least two disjoint,}\\
\;&\quad\;\;\text{ off-diagonal,}\; (E,m_k)\text{-S sub-squares }
\Lam_{L_k}(\ux'),\;\Lam_{L_k}(\ux'')
\},\\
E&= \myset{\text{ both }\Lam'\text{ and }\Lam''\text{ are }E\text{-R}}.
\end{array}\eqno (5.13)$$
Then $\pr{B_k}$ is bounded by
$$
\begin{array}{r}
\pr{B_k \cap C} + \pr{B_k \cap D} + \pr{B_k \cap E}
+ \pr{B_k \cap \bar C \cap \bar D \cap \bar E}
\\
\leq \pr{C} + \pr{ D} + \pr{E} + \pr{B_k \cap \bar C \cap \bar D}.
\end{array}\eqno (5.14)$$
So, it suffices to estimate probabilities
$\pr{C}$, $\pr{ D}$, $\pr{E}$ and
$\pr{B_k \cap \bar C \cap \bar D \cap \bar E}$.

First of all, note that
$$B_k\cap\bar C\cap\bar D\cap\bar E = \emptyset.$$
Indeed, if the potential sample belongs to $\bar C \cap \bar D$, then either of the
sub-squares $\Lam'$, $\Lam''$ contain less than $2n$
$(E,m_k)$-S sub-squares  which are diagonal and $L_k$-D, and at most
one which is off-diagonal. So, the total number of
$(E,m_k)$-S sub-squares of size $L_k$ inside each of the sub-squares
$\Lam', \Lam''$ is bounded by $2n-1 + 1=2n$. In addition, property
$\bar E$ implies that either $\Lam'$ or $\Lam''$ must be $E$-NR.
Witout loss of generality, assume that
$\Lam'$ is
$E$-NR. Then, applying Lemma 4.6 with $K=2n$, we see that $\Lam'$ must be
$(E,m_{k+1}$-NS, which contradicts our hypothesis.

Therefore,
$$\pr{B_k} \leq  \pr{C} + \pr{ D} + \pr{E}.\eqno (5.15)$$

The probability $\pr{C}$ can be estimated with the help of Lemma 4.7. Indeed, set
$$\begin{array}{cl}
C'&=\{\;\Lam' \text{ contains at least }2n\,\,\text{ pair-wise } L_k\text{-D,}\\
\;&\quad\;\;(E,m_k)\text{-S sub-squares } \Lam_k(\uu_i),\;
i=1, \ldots, 2n\;\},\\ \;&\;\\ C''&=\{\;\Lam''\text{ contains at least }2n\,\,\text{
pair-wise } L_k\text{-D,}\\
\;&\quad\;\;(E,m_k)\text{-S sub-squares }
\Lam_k(\uu_i),\;i=1,\ldots, 2n\;\}.\end{array}\eqno(5.16)$$
By virtue of Lemma 4.7,
$$\pr{C'} \leq L_{k+1}^{-n(p-1-\alpha)/\alpha},\;
\;\;\pr{C''} \leq L_{k+1}^{-n(p-1-\alpha)/\alpha}.\eqno(5.17)$$
With our choice (4.1) and with $n \geq \vn$, we get that
$$\frac{n(p - 1 - \alpha)}{ \alpha} >\frac{\vn(\vp -1- \valpha)}{\valpha}
=14 = 2p+2 >2p.$$
Since $C \subset C'\cup C''$, we obtain that
$$\pr{C} < 2 L_{k+1}^{-2p-2}.\eqno(5.18)$$

Next, consider the events
$$\begin{array}{cl}
D'&=\{\;\Lam'\text{ contains at least two disjoint, off-diagonal,}\\
\;&\quad\;\;(E,m_k)\text{-S sub-squares }\Lam_{L_k}(\ux')\text{ and }
\Lam_{L_k}(\ux'')\;\},\\
\;&\;\\
D''&=\{\;\Lam''\text{ contains at least two disjoint, off-diagonal,}\\
\;&\quad\;\;(E,m_k)\text{-S sub-squares }\Lam_{L_k}(\ux')\text{ and }
\Lam_{L_k}(\ux'')\;\}.\end{array}\eqno(5.19)$$
Notice that $D\subset D'\cup D''$, so that $\pr{D}\leq\pr{D'}+\pr{D''}$. Probabilities
$\pr{D'}$ and $\pr{D''}$ are estimated in a similar way, so consider, say, the event
$D'$. Obviously, $D'$ is a union of events $D'(\ux,\uy)$ of the form
$$\begin{array}{cl}
D'(\ux,\uy)&=\{ \Lam' \text{ contains two disjoint, off-diagonal,}\\
\;&\quad\;\;(E,m_k)\text{-S sub-squares, }\Lam_{L_k}(\ux)
\text{ and }\Lam_{L_k}(\uy)\;\},\end{array}\eqno(5.20)$$
and the number of such pairs $(\ux,\uy)$ is bounded by $L_{k+1}^2$. Hence,
$$\pr{D'} \leq L_{k+1}^2 \max_{\ux, \uy}\;\pr{D'(\ux,\uy)}.\eqno (5.21)$$

Now fix a pair of disjoint, off-diagonal sub-squares
$\Lam_{L_k}(\ux)$, $\Lam_{L_k}(\uy)$ and
consider either of them, e.g., $\Lam_{L_k}(\ux)$. By virtue of Lemma 4.3,  if both
coordinate projections $J_1 = \Pi_1 \Lam_{L_k}(\ux)$ and $J_2 = \Pi_2 \Lam_{L_k}(\ux)$
are $(E,m_k)$-NT, then either
$\Lam_{L_k}(\ux)$ is $E$-R or $\Lam_{L_k}(\ux)$
is $(E,m_k)$-NS; the latter is impossible by our hypothesis. Set
$$\tilde J =  \Pi_1 \Lam'_{L_{k+1}} \cup \Pi_1 \Lam''_{L_{k+1}}
\cup \Pi_2 \Lam'_{L_{k+1}} \cup \Pi_2 \Lam''_{L_{k+1}}$$
and consider the event
$$T=\myset{\exists\,\text{ an }(E,m_0+1)\text{-T interval }J_{L_k}
\subset \tilde J }.\eqno (5.22)$$
Then
$$\begin{array}{cl}
\pr{D'(\ux,\uy)}&=\pr{D'(\ux,\uy)\cap T}+\pr{D'(\ux,\uy)\cap\bar T}\\
\;&\leq\pr{ T}+\pr{D'(\ux,\uy)\cap \bar T}.\end{array}\eqno (5.23)$$
Recall that, in the same way as in Lemma 4.5, the value
$m_0$ is chosen so as to
guarantee that properties $(E,m_0+1)$-NT and $E$-NR imply property $(E,m_k)$-NS for any
$k\geq 0$.

By Lemma 3.2 and Corollary 3.1, $\pr{T}\leq L_k^{-q}$. On the other hand, if the
potential sample belongs to $D'(\ux,\uy) \cap
\bar T$, then both $\Lam_{L_k}(\ux)$ and $\Lam_{L_k}(\uy)$ must
be $E$-R. Therefore,
$$\pr{ D'(\ux,\uy) \cap \bar T}
\leq \pr{ \text{both } \Lam_{L_k}(\ux) \text{ and } \Lam_{L_k}(\uy)
\text{ are } E\text{-R}}.$$
Recall that both sub-squares are off-diagonal. Then, by Lemma 4.4, the above probability
is not greater than $L_k^{-q}$. Finally,
$$\pr{D'(\ux,\uy)} \leq L_k^{-q} + L_k^{-q}\eqno (5.24)$$
and
$$\pr{D'} \leq L_{k+1}^2 \, (L_k^{-q} + L_k^{-q})\eqno (5.25)$$
yielding
$$\pr{D} \leq 2 L_{k+1}^2 \, (L_k^{-q} + L_k^{-q}).\eqno (5.26)$$

Finally, probability $\pr{E}$ is estimated again with the help of Lemma 4.4. In fact, the
sub-squares $\Lam'$ and $\Lam''$, being diagonal and $L_k$-D, have both their horizontal
projections disjoint and their vertical projections disjoint. But just one of these
properties would suffice for Lemma 4.4 to be applied:
$$\pr{E} =
\pr{\text{ both }\Lam'\text{ and }\Lam''\text{ are } E\text{-R}}
\leq L_{k+1}^{-q}.\eqno (5.27)$$

Combining bounds (5.15)--(5.27), we see that
$$\begin{array}{cl}
\pr{B_k}&\leq \pr{C} + \pr{D} + \pr{E}\\
\;&\leq 2 L_{k+1}^{-2p}+2 L_{k+1}^2 \, (L_k^{-q} + L_k^{-q})
+ L_{k+1}^{-q}\\
\;&\leq L_{k+1}^{-2p}.\end{array}\eqno (5.28)$$
This proves \DSkone. $\qquad\qed$
\pmn

\myproof{Lemma \ref{DiagAndOffdiag}} {\bf Step 1.} Consider
sub-squares $\Lam' = \Lam_{L_{k+1}}(\ux')$,
$\Lam'' = \Lam_{L_{k+1}}(\ux'')$ and set
$$\tilde J = \Pi_1\Lam' \cup \Pi_2\Lam' \cup \Pi_1\Lam'' \cup \Pi_2\Lam''.
$$
Let $C$ stand for the following event:
$$C = \myset{\;\exists\;\text{ an } (m_k)\text{-T segment } I_{L_k}
\subset \tilde J}.\eqno (5.29)$$
As before, the tunneling property (i.e. delocalisation, or insufficient localisation) for
segments is related to {\it single-particle} spectra. Thus, we can use results of the
single-particle localisation theory (cf. \cite{vDK1}). By Lemma 3.2,
$\pr{C} \leq L_{k}^{-q}$. Next, for
$$
B_k = \myset{\text{ both }\Lam'\text{ and }
\Lam'' \text{ are }(E,m_{k+1})\text{-S }}\eqno (5.30)
$$
we obtain that
$$\pr{ B_k}\leq\pr{C}+\pr{ B_k \cap \bar C}\leq L_{k}^{-q}
+ \pr{ B_k \cap \bar C}.\eqno (5.31)
$$

It now remains to bound probability $\pr{B_k \cap \bar C}$.

\psn{\bf Step 2.} By Lemma 4.3, if one-dimensional projections
of the off-diagonal sub-square (actually, a square) $\Lam''$ are non-tunneling, then
either it is $E$-R, or it is $(E,m_{k+1})$-NS. The latter is impossible for potential
samples in $B_k$ (for both
$\Lam'$ {\it and } $\Lam''$ must be resonant), so $\Lam''$ has
to be $E$-R. Introduce the following event:
$$D = \myset{\text{ both }\Lam' \text{ and }
\Lam'' \text{ are } E\text{-R }}.\eqno (5.32)$$
Since $\Lam' \cap \Lam'' = \emptyset$ and $\Lam''$ is off-diagonal, we can apply Lemma
2.4 and  Lemma 4.4 and write
$$\pr{D} \leq L_k^{-q},\eqno (5.33)$$
so that
$$\pr{ B_k \cap \bar C} \leq \pr{ D }
+ \pr{ B_k  \cap \bar C \cap \bar D}
\leq L_k^{-q} + \pr{ B_k  \cap \bar C \cap \bar D}.\eqno (5.34)$$

\psn
{\bf Step 3.} Assuming now the non-resonance of $\Lam''$ (due to
$\bar D$ and the resonance of $\Lam'$), we see that, due to Lemma
4.6, in order to be resonant, square $\Lam''$ must contain at least $K=2n$ $(E,m_k)$-S
sub-squares $\Lam_{L_k}(\uu)$ of size $L_k$. There are two types of them: diagonal and
off-diagonal. By Lemma 4.7, the probability to have $\geq 2n$ diagonal $L_k$-D, $E$-R
sub-squares $\Lam_{L_k}(\uu_i)$,
$i=1, \ldots, 2n$, is not greater
than $L_k^{n(1+\alpha)} L_k^{-np/2}$. On the other hand, the probability to have both an
off-diagonal (sub-)square $\Lam'$ and an off-diagonal
(sub-)square $\Lam_{L_k}(\uv)$
$E$-R is bounded by $L_k^{-q}$. Combining these two
bounds, we conclude that
$$\pr{B_k \cap \bar C \cap \bar D} \leq L_k^{-q} + L_k^{-q}
= 2 L_k^{-q}.\eqno (5.35)$$

\psn
{\bf Step 4.} With estimates of Steps 1--3 (see Equations (5.31)--(5.35)),
we have that
$$\pr{B_k} \leq L_k^{-q} + L_k^{-q} + 2L_k^{-q}
= 4L_k^{-q} < L_{k+1}^{-2p},\eqno (5.36)$$
with our choice of exponents $p=\vp$, $q=\vq$,
$\alpha=\valpha$. \qquad\qed
\bigskip

{\bf Acknowledgments.} VC thanks The Isaac Newton Institute and Department
of Pure Mathematics and Mathematical Statistics, University of Cambridge, for the
hospitality during visits in 2003, 2004 and 2007. YS thanks D\'{e}partement de Mathematique,
Universit\'{e} de Reims Champagne--Ardenne, for the hospitality during visits in 2003 and
2006, in particular, for the Visiting Professorship in the Spring of 2003 when this work
was initiated. YS thanks IHES, Bures-sur-Yvette, France, for the hospitality during
numerous visits in 2003--2007. YS thanks School of Theoretical Physics, Dublin Institute
for Advanced Studies, for the hospitality during regular visits in 2003--2007. YS thanks
Department of Mathematics, Penn State University, for the hospitality during Visting
Professorship in the Spring, 2004. YS thanks Department of Mathematics, University of
California, Davis, for the hospitality during Visiting Professroship in the Fall of 2005.
YS acknowledges the support provided by the ESF Research Programme RDSES towards research
trips in 2003--2006.

%
%

\end{document}